\documentclass[11pt,letterpaper]{article}
\usepackage[T1]{fontenc}
\usepackage{lmodern}
\usepackage[margin=1in]{geometry}
\usepackage{amsmath,amsthm,amssymb,amsfonts}
\usepackage{microtype,booktabs,float,enumitem,xcolor,tikz}
\floatstyle{ruled}
\newfloat{algorithm}{tbp}{loa}
\floatname{algorithm}{Algorithm}
\definecolor{linkblue}{RGB}{24,59,110}
\definecolor{pine}{RGB}{13,93,87}
\definecolor{ochre}{RGB}{148,72,22}
\usepackage[colorlinks=true,linkcolor=linkblue,citecolor=linkblue,urlcolor=linkblue,
  bookmarksnumbered=true]{hyperref}
\hypersetup{pdftitle={Prophet Secretary for Matroids and XOS Valuations: A Separation at 1-1/e and a 0.6767 Guarantee for I.I.D. Weights},
  pdfauthor={Hossein Esfandiari and MohammadTaghi Hajiaghayi},
  pdfsubject={Online algorithms and prophet inequalities},
  pdfkeywords={matroids, prophet secretary, iid prophet inequalities, XOS valuations}}
\newcommand{\E}{\mathbb E}
\newcommand{\Prb}{\mathbb P}
\newcommand{\OPT}{\operatorname{OPT}}
\newcommand{\ALG}{\operatorname{ALG}}
\newcommand{\cl}{\operatorname{cl}}
\newcommand{\one}{\mathbf 1}

\newtheorem{theorem}{Theorem}
\newtheorem{lemma}[theorem]{Lemma}
\newtheorem{proposition}[theorem]{Proposition}
\newtheorem{corollary}[theorem]{Corollary}
\theoremstyle{definition}

\theoremstyle{remark}

\numberwithin{equation}{section}

\title{Prophet Secretary for Matroids and XOS Valuations\\[3pt]
  \large A Separation at $1-1/e$ and a $0.6767$ Guarantee for I.I.D. Weights}
\author{Hossein Esfandiari\and MohammadTaghi Hajiaghayi}
\date{}
\begin{document}
\maketitle

\begin{abstract}
The matroid prophet-secretary problem asks an online algorithm to select
an independent set of elements whose nonnegative values are drawn
independently and revealed in uniformly random order. Its combinatorial-auction counterpart asks
the algorithm to allocate disjoint bundles to arriving buyers with
independently drawn XOS valuation functions. In both problems the
distributions are known, and the benchmark is the expected offline
optimum. Ehsani, Hajiaghayi, Kesselheim, and Singla (SODA 2018) give
$(1-1/e)$-competitive algorithms for both settings. In this work we show that these
two guarantees have different limits.

For every matroid, we give a randomized
$(1-1/e+\varepsilon_0)$-competitive algorithm, where
$\varepsilon_0=10^{-6}$. The guarantee holds for independent
nonnegative values with finite expected optimum. For explicitly given
finite distributions with rational values and probabilities, the
algorithm has polynomial worst-case running time in the matroid
independence-oracle model. The oracle tests whether a queried set is
independent. The algorithm makes $O(N^4)$ such queries in expectation
and $O(N^4\log(N+2))$ in the worst case, where $N$ is the number of
elements.

We further consider the i.i.d. version of this problem and provide a $0.6767$ competitive ratio for every matroid with i.i.d.\ nonnegative weights from a
known distribution. This guarantee even holds in fixed arrival order (a.k.a. i.i.d. matriod prophet inequalities),
even when the algorithm learns the order only as the elements arrive.

Both matroid proofs measure the cost of acceptance by the rank lost
among unseen elements above a given weight level. With a common law,
these elements form an independent sample at a common rate. The
probability that this sample spans the current element determines a
quantile cutoff that pays for the loss at every level. We determine
the best constant attainable by this one-step comparison. For different
laws, we use a rescaled reference distribution and exploit overlap
between independent optimal bases. Both matroid guarantees admit
polynomial-time implementations for explicitly given finite rational
distributions.

For combinatorial auctions with XOS valuations, however, we show that $1-1/e$ is the optimal
competitive ratio. We give an explicit family of finite instances
showing that no online algorithm can guarantee a larger competitive
ratio than $1-1/e$, even when the buyers'
valuations are i.i.d., every additive
clause has coefficients in $\{0,1\}$, and computation is unrestricted.
Thus the matroid improvement exploits structure absent from general
allocation with XOS valuations.

\end{abstract}

\section{Introduction}

Optimal stopping theory has long informed the design of online auctions.
Hajiaghayi, Kleinberg, and Parkes~\cite{HKP2004} initiated the
connection between secretary problems and online auction design,
and Hajiaghayi, Kleinberg, and Sandholm~\cite{HKS2007} introduced
prophet inequalities as a tool for designing and analyzing online
auction mechanisms.\footnote{These two papers were jointly
	recognized by the 2026 ACM SIGecom Test of Time
	Award~\cite{Kohlhaas2026Award} for their contributions connecting
	optimal stopping with online auction and pricing mechanisms.}
These connections led to a broad study of how distributional
information and random arrival can improve online allocation.

In a prophet inequality problem, the values are drawn independently
from known distributions, and an online algorithm makes irrevocable
decisions as they are revealed. Its performance is compared with a
prophet who sees all realizations before choosing an optimal feasible
solution. The arrival order may be adversarial. In a secretary
problem, the values are fixed and unknown, but the arrival order is
uniformly random. The prophet-secretary model~\cite{EHLM2015}
combines the distributional information of prophet inequalities with
the random arrival order of secretary problems. The distributions are
known in advance, and the arrival order is uniform and independent
of the realized values.

The central difficulty is to use this information while preserving
feasibility for future arrivals. Accepting a valuable element may
prevent the selection of later elements, and allocating an item to
one buyer makes it unavailable to every later buyer. Distributional
information allows an algorithm to estimate these opportunity costs;
random arrival also limits how often an early decision can block a
more valuable later one. We study how the interaction between these
two sources of information depends on the feasibility structure.
Our two settings are matroid selection and combinatorial auctions
with XOS valuations, defined next.

\subsection{Models and results}\label{subsec:models}

\paragraph{Matroid selection.}
The algorithm knows a matroid $M=(E,\mathcal I)$ on $N$ labeled elements
and the distributions of their independent nonnegative weights $X_e$.
On seeing a label and its weight, it irrevocably accepts or rejects
the element. The accepted set must remain independent. In prophet
secretary the arrival order is a uniform permutation independent of the
weights. We will also consider an arbitrary order fixed independently
of the weights and of the algorithm's random bits. The benchmark in
both models is
\[
  F_0=\E\,\OPT(M,X),\qquad
  \OPT(M,x)=\max_{I\in\mathcal I}\sum_{e\in I}x_e.
\]
Write $\ALG$ for the realized online reward. A ratio $c$ means $\E[\ALG]\ge cF_0$. For our computational
bounds the algorithm may query an independence oracle on any subset
of the known ground set.

\begin{theorem}[Independent weights in random order]\label{thm:main}
For every finite matroid and independent nonnegative weights with
$F_0<\infty$, there is a randomized prophet-secretary algorithm with
\[
    \E[\ALG]\ge (1-1/e+\varepsilon_0)F_0,
    \qquad \varepsilon_0=1.14\cdot10^{-6}.
\]
For explicitly given finite rational distributions, it uses $O(N^4)$
independence queries in expectation and $O(N^4\log(N+2))$ in the worst
case. Its arithmetic work and number of random bits are polynomial in
the input length on every execution.
\end{theorem}

When the weights share a law, the next theorem gives a larger guarantee
under a weaker assumption on the order.

\begin{theorem}[I.i.d.\ weights in a fixed order]\label{thm:iid-main}
Let the weights of a known finite matroid be independent draws from a
common known nonnegative distribution of finite mean. There is a
randomized online algorithm that, in every fixed arrival order, obtains
\[
   \E[\ALG]\ge c_0F_0,\qquad
   c_0=\frac{1}{1+\eta_0}=0.6767410571\ldots,
   \qquad \eta_0^2=e^{-1-\eta_0},\quad\eta_0>0.
\]
The algorithm need not know the order in advance.
\end{theorem}

Averaging over a uniform permutation gives the same ratio for prophet
secretary. The theorem also covers a matroid in which some labels
have many substitutes and others have none. Its proof uses symmetry
of the weights alone. The algorithm can be implemented by sampling
and making $O(N^4\log(2N))$ independence queries. In a real-arithmetic
model it only needs evaluations of the common distribution function
and its left limits at the observed weights.
For a finite rational law the same ratio holds with a worst-case polynomial bound on bit complexity (Appendix~\ref{iid:sec:implementation}).

\paragraph{Welfare with XOS buyers.}
There is a fixed set $U$ of items. Each buyer $i$ has a valuation drawn
independently from a known distribution. When buyer $i$ arrives, the
algorithm observes her valuation $v_i$ and assigns her a bundle $S_i$
of unallocated items. Her valuation is \emph{XOS} if it is the maximum
of nonnegative additive clauses, that is,
$v_i(S)=\max_{\ell}\sum_{j\in S}w_{i\ell j}$ for $w_{i\ell j}\ge0$.
The benchmark $\OPT$ is the maximum welfare $\sum_i v_i(S_i)$ over
pairwise disjoint bundles for the realized valuations.

\begin{theorem}[Sharpness for XOS welfare]\label{thm:xos-upper}
For every $\xi>0$ there is a finite instance with i.i.d.\ buyers for
which every randomized online algorithm satisfies
\[
    \E[\ALG]<(1-1/e+\xi)\E[\OPT].
\]
All additive clauses have coefficients in $\{0,1\}$. The bound holds
with public menus of clauses and unlimited computation, even if the
order of buyer identities is announced in advance.
\end{theorem}

Together with the algorithm of Ehsani et al.~\cite{EHKS2018}, this
identifies the optimal guarantee for XOS welfare. The upper bound also
follows from the i.i.d.\ coverage-welfare construction of Udwani and
Wajc~\cite{UW2026} by interchanging the roles of items and buyers.
Section~\ref{sec:xos-upper} gives the reduction and a direct proof.

\paragraph{Relation to the earlier version.}
Theorem~\ref{thm:main} of the first arXiv version provides a similar competitive ratio with $\varepsilon_0=1.01\cdot10^{-6}$ for matroid prophet secretary
with the same oracle complexity. The main addition here is Theorem~\ref{thm:iid-main} for i.i.d. distributions, with an improved competitive ratio. The previous version also established the
XOS upper bound. In addition to that, here, we show that the XOS upper bound with
additive tolerance $\xi$ can be witnessed on
$O(\xi^{-4}\log(1/\xi))$ items (Corollary~\ref{cor:xos-items}).

\subsection{Approach and techniques}\label{subsec:overview}

\paragraph{The common accounting identity.}
For a nonnegative weight vector $x$, put $S_t=\{e:x_e>t\}$ and let
$r(S)$ denote the rank of $S$. A maximum-weight basis found by greedy
contains exactly $r(S_t)$ elements of $S_t$ at every height $t$.
Consequently,
\begin{equation}\label{eq:intro-layers}
                    \OPT(M,x)=\int_0^\infty r(S_t)\,dt .
\end{equation}
To see what one online decision costs at a given height, fix the arriving
nonloop $e$, and let $A$ consist of the unseen elements above that
height. If $A$ spans $e$, taking $e$ consumes one unit of the rank
available to $A$. If $A$ does not span $e$, taking $e$ costs $A$
nothing, but passing over $e$ can lose a unit of rank when its own
weight exceeds $t$. Both matroid algorithms price these two possibilities.
The common-law assumption changes what can be learned from a single
sampling experiment.

\paragraph{One quantile cutoff for every height.}
Suppose the common upper-tail probability at height $t$ is $v$.
In the residual matroid, the unseen elements above $t$ form a set
obtained by retaining each unseen element independently with probability
$v$. Write $S_e(v)$ for the probability that this set fails to span
$e$. An upper quantile $q$ specifies a threshold. The rule takes $e$
on the best $q$ fraction of its draws, breaking ties at atoms by an
independent random choice. At any height with tail probability $v$ this
rule earns $\min\{q,v\}$ in expected count. Its expected cost to the
remaining rank is
\[
                  vS_e(v)+q(1-S_e(v)).
\]
The first term is the expected cost of deletion. The second is the extra
cost of contracting $e$ when the rule accepts. In particular, the
decision pays for itself exactly at $v=q$.

The same cutoff must also pay for the narrow bands $v<q$ and the wide
bands $v>q$. For $v<q$ there is a loss only when the unseen sample spans $e$.
For $v>q$ there is a loss only when it fails to do so. This is where the shape of
$S_e$ matters. Calibrate $q$ by $S_e(q)=e^{-a}$. A product-measure
inequality bounds $S_e(v)$ from below by $e^{-av/q}$ for $v<q$ and
from above by the same curve for $v>q$. Thus the excess cost per unit
of reward is at most $a$ on narrow bands and at most
$\max_{x\ge1}(x-1)e^{-ax}=e^{-1-a}/a$ on wide bands. Equating the two
gives exactly $a=\eta_0$ in Theorem~\ref{thm:iid-main}. The bound holds
after fixing both the history and the next label, which explains why
this proof needs no randomness in the arrival order. A sequence of
parallel elements makes the one-step comparison sharp. The optimal
rank-one policy can do better by compensating for a loss at one step
with a gain at another.

\paragraph{Beyond the one-step comparison.}
For i.i.d.\ weights in random order, the distinction between a local
guarantee and the optimal ratio can be made precise. Follow an arbitrary
policy through its sequence of residual matroids. If it accepts a label
with conditional probability $q$ in a given minor, replacing that decision
by a top-$q$ cutoff sends the same probability mass to each child and
earns at least as much in every upper band. Thus a policy can be described
by a flow on labeled minors. In fact a single flow attains the
worst-law ratio of the matroid simultaneously for all common laws.
If some finite set of band targets were incompatible, a separating
hyperplane would supply nonnegative weights for the bands. Those
weights are the increments of a finite-support common law on which
every policy misses its target. Appendix~\ref{iid:sec:flows} proves
this characterization and a companion identity that compares the optimal
online price with the prophet's realized replacement cost on the same
random restriction.

\paragraph{Pricing a decision when the laws differ.}
With arbitrary marginals, the unseen set at height $t$ samples
different labels at different rates. There is no common quantile at
which to apply the preceding comparison. We work directly with the
weighted optimum and use the randomness of the next label.
For a nonloop $e$, fix the other weights and let $\tau_e$ be the
difference between their optima after deleting and contracting $e$.
Deleting $e$ from the original instance loses $(X_e-\tau_e)^+$, where
$u^+=\max\{u,0\}$. The sum of these losses in expectation, divided
by the number of remaining labels, is the cost of rejecting the next
arrival.

A large expected deletion loss can arise in two ways.
An element might rarely draw a very large weight, or it might
regularly beat its replacement by a substantial margin. To distinguish them, draw
its weight a second time, keeping the other weights fixed. The
quantity $(\min\{X_e,X'_e\}-\tau_e)^+$ counts the part of the
surplus that survives the second draw. Summing its expectation over labels gives the
\emph{stable surplus}. Roughly speaking, rare high offers favor a
posted price, whereas persistent membership in an optimum favors a
basis-exchange rule. To make this distinction quantitative, we first
choose the distribution against which to measure the two surpluses.

Starting from the actual upper tails, multiply every tail probability by
a common scale factor $q$ near one, capping the result at one. Let $F_q$
be the expected optimum of this reference instance. The potential
is $(1-\theta q)F_q$, with $\theta$ slightly below $1/e$.
Increasing $q$ makes the reference optimum larger and its coefficient
smaller. The rate at which $F_q$ grows is closely related to deletion
surplus. At a given height, increasing one label's probability changes
expected rank precisely when the other active labels fail to span it.
Integrating over heights relates the derivative of $F_q$ to the sum of
the deletion losses. The stable surplus bounds the contribution of tails that
reach the cap. Concavity then allows the algorithm to choose $q$ without
sacrificing the potential inherited from earlier decisions.

\emph{Small stable surplus.}
The algorithm posts a price proportional to the expected difference
between deleting and contracting the next label. Charging the price
only to elements of the optimal basis loses too much. An element
outside that basis can still clear its price, and a rule processing
one arrival can collect its surplus. To quantify this contribution we
cap the reference weights at scaled contraction prices and apply a
tangent inequality for the expected rank under Poisson
rounding~\cite{DRY2011}. After integrating over heights, the extra
surplus covers the expected deletion loss. The rule processes one
arrival and repeats in the resulting minor.

\emph{Large stable surplus.}
Two independent reference optima then have substantial overlap.
To measure the relevant overlap, count a shared element at height $t$
only when it belongs to both optima and both its weights exceed $t$.
Integrating this expected count gives a weighted overlap $Q$.
Section~\ref{sec:stable-overlap} lower-bounds $Q$ by comparing greedy and
reverse-greedy bases of each level set. The algorithm can spend this
overlap in the last part of the arrival sequence. It maintains virtual
bases containing all previously accepted labels, updating each basis by
an injective exchange whenever a new label is accepted. Early on it
randomly thins acceptances to control the rate at which an exchange can
displace any particular element. In a short final interval it removes
this thinning. The corresponding loss in the baseline guarantee is
quadratic in the length of that interval, whereas the gain from overlap
is linear. Near the endpoints of the reference search, the baseline
continuation has enough slack to pay for the final interval even at zero overlap. These comparisons cover all
branches of the algorithm.

\paragraph{Why the XOS upper bound survives identical distributions.}
The auction construction has a public collection of partitions of the
items, with one partition for every possible set of distinct buyer
types. If the $n$ realized types are distinct, the prophet can choose
their common partition and give each buyer a different part, attaining
full welfare. An online allocation fixes a supporting clause for each
bundle as soon as it is assigned. Each such clause points to a partition
indexed by just $n$ types, so with a sufficiently large type space a
later buyer is unlikely to be able to use the same partition. The
construction makes one part from each of $n$ distinct partitions cover
exactly a $1-(1-1/n)^n$ fraction of the items. This counting identity
limits the total welfare even when all valuation types and their
probabilities are public.

\subsection{Related work}

\paragraph{Independent distributions and random arrival.}
The classical prophet inequality compares an online stopping rule with
the expected maximum of independent nonnegative random variables~\cite{ks77}.
Its connections to online auctions and pricing were developed
in~\cite{HKP2004,HKS2007}.\footnote{These two papers received the
2026 ACM SIGecom Test of Time Award~\cite{Kohlhaas2026Award}.} Esfandiari et al.~\cite{EHLM2015} introduced
the prophet-secretary model. In an arbitrary fixed order, the matroid
inequality of Kleinberg and Weinberg~\cite{KW2012} and the posted-price
mechanism of Feldman, Gravin, and Lucier~\cite{FGL2015} give a factor
of $1/2$ for matroid selection and XOS welfare, respectively.
Ehsani et al.~\cite{EHKS2018} showed that random arrival raises both
guarantees to $1-1/e$. For a single choice, Azar, Chiplunkar, and
Kaplan~\cite{ACK2018} broke this barrier, followed by
Correa, Saona, and Ziliotto~\cite{CSZ2021}. Chen et al.~\cite{CHLT2025}
obtained $0.688$ for a single choice and $0.641$ for prophet-secretary
matching.

\paragraph{Identical distributions.}
In rank one the sharp constant over all horizons is
$0.745440\ldots$. Hill and Kertz~\cite{hk82} obtained the sharp
finite-horizon constants, and Kertz~\cite{kertz} determined their
limit. Abolhassani et al.~\cite{aee} gave a threshold rule attaining
$0.738$ asymptotically. Correa et al.~\cite{cfhov} matched the limiting
upper bound with a guarantee for every horizon. For $k$ choices,
Brustle, Perez-Salazar, and Verdugo~\cite{bpv} characterized the
finite-horizon i.i.d. ratio by a linear program. A cardinality
constraint is invariant under relabeling. On a general matroid,
exchanging two labels can change feasibility even when their weights
have the same law. In the related random-assignment secretary problem~\cite{soto}, an unknown
list of weights is assigned uniformly to the labels. Conditioning i.i.d.\ weights on their unordered multiset gives this assignment. Our algorithm also uses the common law to compute weight quantiles.

\paragraph{Contention resolution and basis exchange.}
An online contention-resolution scheme selects a feasible subset from
a random set of active elements~\cite{FSZ2016}. Lee and
Singla~\cite{LS2018} obtained the tight factor $1-1/e$ for matroids
under random arrival against the ex-ante relaxation. The distinction
between benchmarks already matters in rank one. If each of $n$
elements independently has value one with probability $1/n$ and zero
otherwise, the ex-ante value is one, whereas the expected optimum is
$1-(1-1/n)^n$. Our matroid guarantees use the latter benchmark.
Adamczyk and W{\l}odarczyk~\cite{AW2018} used basis-exchange maps
of the kind proved by Brualdi~\cite{Brualdi1969} to maintain
feasible sets under random arrivals. Our heterogeneous algorithm
applies exchanges to bases drawn from a reference distribution and
uses the overlap of two such optima to bound the gain over $1-1/e$.
The rank inequality in its analysis uses the concavity of
Poisson-rounded matroid rank~\cite{DRY2011}.

\paragraph{Information about weights.}
In the matroid secretary problem~\cite{BIK2007}, the weights are
fixed and unknown. Recent algorithms give a constant guarantee in
this model~\cite{Singla2026,Huang2026,ABHHMS2026}. The single-sample
prophet model gives one independent sample from each law. Azar, Kleinberg,
and Weinberg studied its connection to secretary algorithms~\cite{AKW2014}. For a fixed arrival order independent
of samples, realizations, and private randomness, Abdi
et al.~\cite{ABHM2026,ABHHMS2026} obtained the sharp factor $1/2$
on every matroid. Our algorithms use known distributions.

\paragraph{Online welfare.}
In the item-arrival model with coverage valuations, Kapralov, Post,
and Vondr\'ak~\cite{KPV2013} proved that greedy achieves $1-1/e$ on
i.i.d. arrivals and that a polynomial-time algorithm improving this
factor would imply $\mathrm{NP}=\mathrm{RP}$. Udwani and
Wajc~\cite{UW2026} proved the same upper bound for unrestricted
computation even when the common distribution is known. Chawla and
Dang~\cite[Appendix~D]{CD2026} used a coverage-to-XOS transformation
in the free-disposal model. Section~\ref{subsec:xos-coverage}
explains how it carries the information-theoretic bound over to
irrevocable buyer-arrival allocation.

\paragraph{Organization.}
After the common matroid preliminaries, Section~\ref{iid:sec:band} gives
the i.i.d.\ algorithm and proves the sharp bound on its one-step analysis.
For general marginals, Sections~\ref{sec:structure}
and~\ref{sec:controller} prove the threshold and exchange bounds.
Sections~\ref{sec:combined-estimation}--\ref{sec:policy} use these bounds
in the online algorithm. Section~\ref{sec:xos-upper} gives the auction
construction. Appendices~\ref{iid:sec:implementation}
and~\ref{iid:sec:flows} treat implementation and optimal policies for
i.i.d.\ weights. Appendices~\ref{sec:combined-implementation}
and~\ref{sec:parameters} give the bit complexity and the numerical
certificate for general marginals.

\section{Matroid preliminaries}\label{sec:preliminaries}

\paragraph{Matroids and residual instances.}
We use the matroid and prophet-secretary models of
Section~\ref{subsec:models}. For $M=(E,\mathcal I)$, the \emph{rank}
of $S\subseteq E$ is
\[
 r_M(S)=\max\{|I|:I\subseteq S,\ I\in\mathcal I\}.
\]
A basis is a maximal independent set. Write $\mathcal B(M)$ for the set of bases, all of size $r_M(E)$. The \emph{closure} of $S$ is
$\cl_M(S)=\{e\in E:r_M(S\cup\{e\})=r_M(S)\}$.
We say $S$ \emph{spans} $e$ if $e\in\cl_M(S)$ and spans $T$ if
$T\subseteq\cl_M(S)$. A \emph{flat} is a set equal to its closure.
An element is a \emph{loop} if its singleton is dependent and a
\emph{coloop} if it belongs to every basis. Two distinct nonloops
are \emph{parallel} if their pair has rank one. Subscripts are
omitted when the matroid is clear.

The \emph{restriction} to $R\subseteq E$ is
$M|_R=(R,\{I\in\mathcal I:I\subseteq R\})$, and the
\emph{deletion} of $R$ is $M\setminus R=M|_{E\setminus R}$.
The \emph{contraction} $M/R$ has ground set $E\setminus R$ and
rank function
\[
 r_{M/R}(S)=r_M(S\cup R)-r_M(R).
\]
This definition also applies when $R$ is dependent. If $R$ is
independent, $S$ is independent in $M/R$ exactly when
$S\cup R$ is independent in $M$. For a single element, write
$M\setminus e$ and $M/e$. A matroid obtained by deletions and
contractions is a \emph{minor}. A set $J$ disjoint from $T$ is
\emph{independent over $T$} if
$r_M(T\cup J)=r_M(T)+|J|$.

A \emph{circuit} is an inclusionwise minimal dependent set. If
$B$ is a basis and $e\notin B$, the unique circuit in
$B\cup\{e\}$ is the \emph{fundamental circuit} of $e$ with
respect to $B$. We use the shorthand
$B-f+e=(B\setminus\{f\})\cup\{e\}$.

At a residual state, the accepted independent set is contracted
and rejected elements are deleted. We discard loops. Independence of the weights
allows us to use the original distribution on each surviving label. To implement
a matroid independence query in the residual minor, adjoin the
already contracted independent set and make one query to the
original matroid independence oracle.

\paragraph{Weighted bases.}
We fix a total order of the labels throughout. The greedy algorithm
scans elements in decreasing order of weight, using this label order
to break ties, and adds an element whenever independence is preserved.
It chooses a canonical maximum-weight basis, denoted $B(y)$.
Zero-weight elements are used to complete the basis.
A \emph{product distribution} is a joint distribution with mutually
independent coordinates. For a value vector $Y$ with a product
distribution, write $F(M,Y)=\E\OPT(M,Y)$, omitting the matroid and
distribution when fixed. The distributions may have atoms, including an atom at zero. We write
$u^+=\max\{u,0\}$, use $\one\{A\}$ for the indicator of an
event $A$, and put $[k]=\{1,\ldots,k\}$.

\section{Identically distributed weights}\label{iid:sec:band}

Suppose that the weights are independent draws from a common distribution
$F$ on $[0,\infty)$ of finite mean. Fix an arrival order independently of
the draws and the algorithm's randomness. We now prove Theorem~\ref{thm:iid-main}. At each step $N$ denotes the
minor on the remaining labels, obtained by contracting accepted elements
and deleting rejected elements. We write $P_N(F)$ for its expected
offline optimum.

Randomly break ties within each atom of $F$. More precisely, when $e$ has
weight $x$, draw its \emph{upper quantile} $U_e$ uniformly from
$[1-F(x),1-F(x-)]$. These quantiles are independent and uniform on
$[0,1]$, with the smallest quantiles corresponding to the largest weights.
For $0\le v\le1$, let $B_v$ be the elements with quantiles at most $v$.
Equivalently, sample each label independently with probability $v$. Define
\[
 R_N(v)=\E r_N(B_v),\qquad
 S_e^N(v)=\Pr\bigl(e\notin\cl_N(B_v\setminus\{e\})\bigr).
\]
Here $B_v$ is sampled on $E(N)$, and $r_N$ is the rank function of $N$.
The function $S_e^N$ measures how hard it is to replace $e$ at a given
weight level. In particular, it depends on the unseen labels, but has
no dependence on their order.

\paragraph{The rule.}
Fix a parameter $a>0$. Reject loops and accept coloops. For any other
arrival $e$, accept if $S_e^N(U_e)\ge e^{-a}$. Equivalently, accept the
best $q$ fraction of the draws, where $q$ is the solution to
$S_e^N(q)=e^{-a}$. The solution is unique since $S_e^N$ is a nonincreasing
polynomial, with $S_e^N(0)=1$ and $S_e^N(1)=0$. Its values lie strictly
between $0$ and $1$ in the interior. The sampled implementation in
Appendix~\ref{iid:sec:implementation} needs only independence queries.

\begin{theorem}[One cutoff for all weight levels]\label{iid:thm:band}
For every $v\in[0,1]$, the expected number of accepted labels in
$B_v$ is at least $c(a)R_M(v)$, where
\[
 c(a)=\frac{1}{1+\max\{a,e^{-1-a}/a\}}.
\]
Consequently the rule earns at least $c(a)P_M(F)$ in every fixed order.
The best choice of $a$ is the unique positive solution $\eta_0$ to
$\eta_0^2=e^{-1-\eta_0}$, giving $c_0=(1+\eta_0)^{-1}=0.6767410571\ldots$.
\end{theorem}

The proof uses one property of product measures. It controls the
probability that $e$ increases rank at \emph{all} sampling rates from
its value at the cutoff.

\begin{lemma}\label{iid:lem:shape}
If $e$ is neither a loop nor a coloop, let $S(v)=S_e^N(v)$ and
$\tau(v)=-\log(1-v)$. Both $-\log S(v)/\tau(v)$ and
$-\log S(v)/v$ are nondecreasing on $(0,1)$.
\end{lemma}

\begin{proof}
Record an absent label by a $1$, so the bits have bias $s=1-v$.
The event that $e$ is unspanned is increasing in these bits and has
probability $h(s)=S(1-s)$. The biased edge-isoperimetric inequality on
the discrete cube~\cite{ekl} gives
\[
                 -h(s)\log h(s)\le -s\log(s)\,h'(s).
\]
Indeed, the entropy of the event indicator is $-h\log h$.
By tensorization it is bounded by the sum of the expected conditional
entropies obtained by resampling each bit in turn. The conditional
entropy of the indicator in a given coordinate is $-s\log s$ when it is pivotal and zero otherwise.
Differentiating the product probability shows that the sum of the
pivotal probabilities is $h'(s)$.

The displayed inequality is equivalent to
$(\log h(s)/\log s)'\le0$. Replacing $s$ by $1-v$ gives the first
claim. The second follows on multiplying by $\tau(v)/v$, the increasing
average of $1/(1-t)$ over $0\le t\le v$.
\end{proof}

\begin{figure}[tbp]
\centering
\begin{tikzpicture}[x=5.3cm,y=2.45cm,font=\small]
 \pgfmathsetmacro{\cut}{sqrt(1-exp(-0.4776700622632))}
 \draw[->] (0,0)--(1.08,0) node[right] {$v$};
 \draw[->] (0,0)--(0,1.12);
 \node[left] at (0,1) {$1$}; \node[below left] at (0,0) {$0$};
 \node[below] at (1,0) {$1$};
 \draw[densely dotted,gray] (\cut,0)--(\cut,{1-\cut*\cut});
 \node[below] at (\cut,0) {$q$};
 \draw[pine,very thick,domain=0:1,samples=80] plot(\x,{1-\x*\x});
 \draw[ochre,thick,dashed,domain=0:1,samples=80] plot(\x,{exp(-0.4776700622632*\x/\cut)});
 \fill (\cut,{1-\cut*\cut}) circle[radius=1.5pt];
 \draw[pine,very thick] (1.20,.83)--(1.32,.83) node[right,black] {$S(v)=1-v^2$};
 \draw[ochre,thick,dashed] (1.20,.58)--(1.32,.58) node[right,black] {$e^{-\eta_0v/q}$};
\end{tikzpicture}
\caption{In a three-element circuit both other labels must be sampled
to span the current one, so $S(v)=1-v^2$. The exponential through the
cutoff bounds the probability from below for $v<q$ and from above for
$v>q$. Lemma~\ref{iid:lem:shape} gives these same two bounds for every
matroid.}\label{iid:fig:cross}
\end{figure}
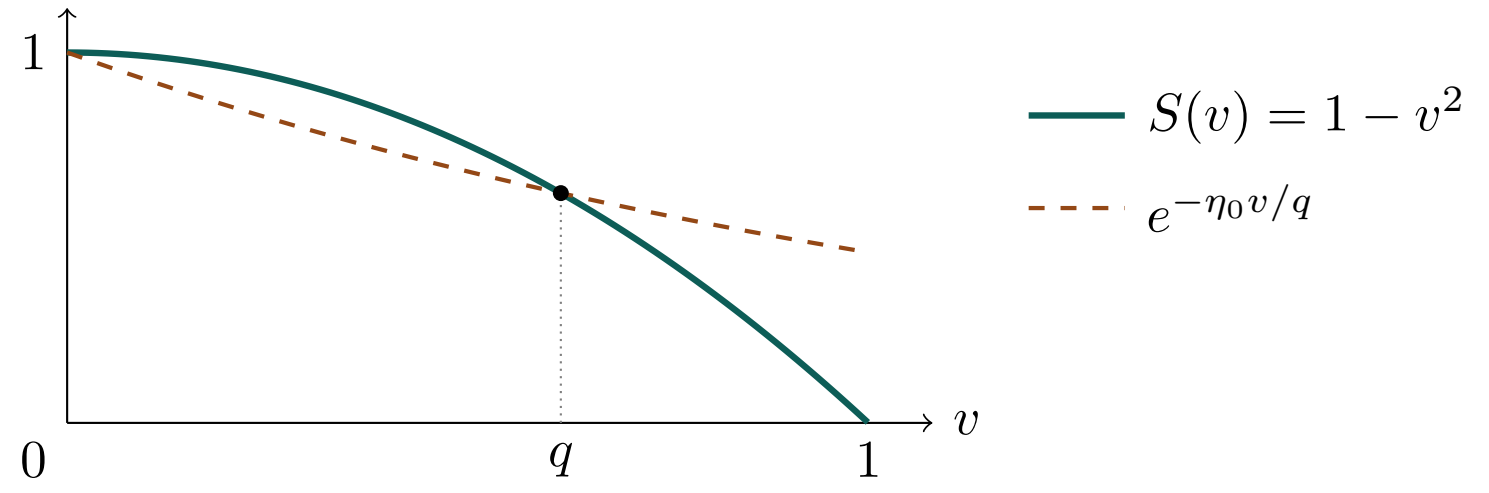

\begin{proof}[Proof of Theorem~\ref{iid:thm:band}]
Fix $v>0$. Condition on the history before the next weight is revealed,
and let $e$ be the next label. At a nonloop, write $S=S_e^N$ and
$N^-=N\setminus e$, $N^+=N/e$. The expected rank increases by $vS(v)$
on adding $e$ to a Bernoulli sample of the other labels. Also,
$r_{N/e}(A)=r_N(A\cup\{e\})-1$, so contracting $e$ costs the sample
one unit of rank precisely when it spans $e$. Thus
\[
 R_N(v)-R_{N^-}(v)=vS(v),\qquad
 R_{N^-}(v)-R_{N^+}(v)=1-S(v).
\]
With acceptance probability $q$, the expected drop in the potential
$R_N(v)$ is
\begin{equation}
 D(v)=q+(v-q)S(v).                                      \label{iid:eq:drop}
\end{equation}
An upper-quantile cutoff of size $q$ earns $\min(v,q)$ in the band.
Thus the cost of using $q$ instead of the best cutoff for this
particular band is exactly
\[
 D(v)-\min(v,q)=
 \begin{cases}
 (q-v)(1-S(v)),&v\le q,\\
 (v-q)S(v),&v\ge q.
 \end{cases}
\]
In the first case we sometimes accept below the band and block a future
element in it. In the second we sometimes reject an element in the band
that the other band elements cannot replace.

For an element that is neither a loop nor a coloop, set $x=v/q$.
As illustrated in Figure~\ref{iid:fig:cross}, Lemma~\ref{iid:lem:shape} and $S(q)=e^{-a}$ give $S(v)\ge e^{-ax}$
for $x\le1$ and $S(v)\le e^{-ax}$ for $x\ge1$, with the endpoint
values obtained by continuity. In the first case
\[
 \frac{D(v)}{v}-1\le (x^{-1}-1)(1-e^{-ax})\le a.
\]
In the second case
\[
 \frac{D(v)}q-1\le (x-1)e^{-ax}\le e^{-1-a}/a.
\]
The last maximum is attained at $x=1+1/a$. In either case the expected
reward in the band pays for $c(a)D(v)$. At a loop both quantities
vanish, and at a coloop the reward and the rank drop are both $v$.
Summing over arrivals telescopes the rank potential from $R_M(v)$ to
zero. The assertion for $v=0$ is immediate.

To recover weight, put $p_t=\Pr(X>t)$. Matroid greedy in descending
weight order takes a basis of every upper level set. Hence
$P_M(F)=\int_0^\infty R_M(p_t)\,dt$. The algorithm's expected weight
is the same integral of its expected band counts. Integrating proves
the guarantee. Finally, $a$ increases and $e^{-1-a}/a$ decreases from
infinity to zero, so their unique intersection maximizes $c(a)$.
\end{proof}

\begin{corollary}[A threshold in the original weights]\label{iid:cor:price}
Another $c_0$-competitive rule accepts a nonloop $e$ exactly when
\[
 X_e\ \ge\ c_0\bigl(P_{N\setminus e}(F)-P_{N/e}(F)\bigr).
\]
\end{corollary}
\begin{proof}
Integrating the one-step inequality in the proof of
Theorem~\ref{iid:thm:band} shows that the expected current reward plus
$c_0$ times the expected prophet value in the child minor is at least
$c_0P_N(F)$. For each realized $X_e$, the rule in the corollary
maximizes this sum over the two choices. Applying the inequality at each
arrival proves the claim.
\end{proof}

This price also connects the two matroid bounds. Let $\tau_e$ be the
difference of the realized optima in $N\setminus e$ and $N/e$ on
the remaining weights. The prophet's gain over deletion is
$\E(X_e-\tau_e)^+$; the best decision credited with $c_0$ times the
child's prophet value gains $\E(X_e-c_0\E\tau_e)^+$ over its deletion
credit. Corollary~\ref{iid:cor:price} and its proof show that, for a
common law, the latter gain pays at least $c_0$ times the former for
\emph{each} next label. With different marginals we will obtain a
surplus comparison by averaging over the next label under a suitably
chosen reference law (Section~\ref{sec:policy}).

\subsection{How much can a one-step comparison prove?}\label{iid:sec:sharpness}

The quantitative loss in Theorem~\ref{iid:thm:band} is sharp for the
local comparison used in its proof. This statement already holds in
rank one, where the optimal stopping rule attains the larger Hill--Kertz
constant~\cite{cfhov}. The example identifies where a stronger argument
must use the continuation of the policy.

\begin{proposition}\label{iid:prop:local-sharpness}
Suppose a rule, possibly randomized, earns at every arrival and every
quantile level $v$ at least a fraction $c$ of the expected one-step
drop in $R_N(v)$, on every matroid with a common weight law. Then
$c\le c_0$. The same upper bound holds for a rule required only to
cover the one-step drop in $P_N(F)$, even when the rule can depend on $F$.
\end{proposition}
\begin{proof}
For the band comparison, consider $m+1$ parallel elements. Write
$q_m$ for the marginal probability of taking the first element. At
level $v$ the first-step reward is at most $\min(q_m,v)$ for any
decision rule. By~\eqref{iid:eq:drop}, the expected drop is
$D_m(v)=q_m+(v-q_m)(1-v)^m$. For $0<c<1$, put $d=1/c-1$.
A positive guarantee requires $q_m>0$. Taking $v\downarrow0$ gives
$1+mq_m=\lim_{v\downarrow0}D_m(v)/v\le1+d$.
Now set $v=(1+d)/m$ for large $m$, so $v>q_m$. The local inequality
would give
\[
 d\ge D_m(v)/q_m-1=(v/q_m-1)(1-v)^m
   \ge \frac{1}{d}\left(1-\frac{1+d}{m}\right)^m .
\]
Passing to the limit yields $d^2\ge e^{-1-d}$, equivalently $c\le c_0$.
The case $c=1$ follows by monotonicity.

For the expected-weight comparison, suppose $c_0<c\le1$. Let $M_k$
denote the maximum of the $k$ future draws on $k+1$ parallel elements
and let $X$ be the first draw. Crediting the child with $c$ times its
prophet value, the best decision compares $X$ with $c\E M_k$.
Subtracting the deletion credit $c\E M_k$ from both sides, the
required inequality becomes
\begin{equation}
 \E(X-c\E M_k)_+\ge c\E(X-M_k)_+.                 \label{iid:eq:priced-sharp}
\end{equation}
With $d=1/c-1$ and $a=1+d$, take the three-point law on
$\{0,1,H_k\}$ specified by
\[
 \Pr(X\ge1)=a/k,\qquad \Pr(X=H_k) = k^{-2},\qquad
 H_k-1=\frac{d+(1-a/k)^k}{1-(1-k^{-2})^k}.
\]
For all large $k$ these are valid probabilities and $H_k>1$.
The choice of $H_k$ makes $c\E M_k=1$ and
$(H_k-1)/k\to b=d+e^{-a}$. Multiply~\eqref{iid:eq:priced-sharp}
by $k$. Its left side tends to $b$. On its right use
$\E(X-M_k)_+=\int_0^\infty (1-F(t))F(t)^k\,dt$. The contributions
from $[0,1)$ and $[1,H_k)$ after multiplying by $k$ tend to $ae^{-a}$
and $b$, respectively. Thus $b\ge c(ae^{-a}+b)$ would be necessary,
or $d^2\ge e^{-1-d}$. This contradicts $c>c_0$.
\end{proof}

In rank one, a stopping rule attains the larger Hill--Kertz constant by
balancing gains across arrivals. Whether the same $0.74544\ldots$
guarantee holds on every matroid in random order remains open.
Appendix~\ref{iid:sec:flows} characterizes the optimum in that model
by allowing precisely this kind of compensation.

\section{Reference distributions and structural inequalities}
\label{sec:structure}

The algorithm compares product distributions obtained by multiplying
every upper tail by the same factor. This section establishes the two
inequalities it needs. The first converts surplus that survives a second
draw into overlap between two optimum bases
(Lemma~\ref{lem:stable-overlap}). The second lower-bounds the surplus
available to an online threshold rule when little of the deletion
surplus survives (Corollary~\ref{cor:structural-surplus}).

\paragraph{Information and computation.}
The algorithm knows the ground set, the matroid through its
independence oracle, and the labeled input distributions. At an
arrival it observes the label and value and must decide immediately.
The algorithm draws its reference samples using fresh randomness.
When it constructs a candidate for the currently arriving label, it
may fix that coordinate at the observed value and sample all other
coordinates independently from their reference distributions.

For the computational part of Theorem~\ref{thm:main}, each distribution is an explicit
finite list of rational values and probabilities; let $L$ be their
total binary encoding length. We first analyze exact random draws.
Appendix~\ref{sec:combined-implementation} replaces them by finite
procedures and proves a worst-case polynomial bound on arithmetic
operations and random bits. 

\paragraph{Symbolic parameters.}
Fix positive rational constants $\delta,\theta,\varepsilon,\eta,\rho,s$,
with $\delta$ dyadic, meaning an integer divided by a power of two. Define
\begin{equation}\label{eq:parameter-definitions}
 J=[q_-,q_+]=[1-4\delta,1+4\delta],\qquad
 a(q)=1-\theta q,\qquad
 d_*(q)=\frac{1-a(q)}{a(q)}.
\end{equation}
The potential is $a(q)F_q$, where $F_q$ is the expected reference
optimum defined below. The algorithm uses $\eta$ to decide whether to
invoke the basis-exchange continuation. Its last $\delta$ units of time
will play a different role from the rest. The estimates have accuracy
parameter $\varepsilon$ and error budget $\rho$. The constant $s$ enters
only in the rank inequality of Lemma~\ref{lem:core-poisson-tangent}.
We use the following bounds:
\begin{equation}\label{eq:basic-parameters}
 \begin{gathered}
 0<\varepsilon<\delta\le\tfrac18,\qquad
 \varepsilon\le\tfrac1{10},\qquad
 0<\theta<1,\quad 0<\rho<1,\quad s>0,\quad \eta>10\varepsilon,\\
 \tfrac12\le a(q)\le\tfrac34\quad(q\in J),\qquad
 \bigl(\max_{q\in J}a(q)\bigr)q_+\le1,\qquad
 d_*(q)-20\varepsilon<1\quad(q\in J).
 \end{gathered}
\end{equation}
The proof of Theorem~\ref{thm:main} will use three further scalar inequalities,
stated in Section~\ref{sec:policy}. Their meaning will be clear
after the two possible continuations have been analyzed.
Appendix~\ref{sec:parameters} gives a single choice satisfying all
of them.

\subsection{Deletion surplus and reference distributions}
\label{sec:reference-foundations}

Let $M$ be a loopless matroid, let $Y=(Y_e)_{e\in E}$ have independent
nonnegative coordinates, and assume
$F=\E\OPT(M,Y)<\infty$. Fix a total order on the labels and use it to
break all greedy ties. Let $B(Y)$ be the resulting maximum-weight basis;
zero-weight elements complete an optimum independent set to a basis.

For each label $e$, define its exclusion threshold by
\[
 \tau_e(Y_{-e})=
 \OPT(M\setminus e,Y)-\OPT(M/e,Y).
\]
This threshold depends only on the other values and is zero for a
coloop. Separating an optimum according to whether it contains $e$
gives
\begin{equation}\label{eq:deletion-contraction-identities}
 \begin{aligned}
 \OPT(M,Y)-\OPT(M\setminus e,Y)&=(Y_e-\tau_e)^+,\\
 \OPT(M,Y)-\OPT(M/e,Y)&=\max\{Y_e,\tau_e\}.
 \end{aligned}
\end{equation}
Write
\[
 \begin{aligned}
 m_e&=\E\tau_e,&
 d_e&=\E(Y_e-\tau_e)^+,& D&=\sum_e d_e,\\
 H&=\sum_e\E(\min\{Y_e,Y'_e\}-\tau_e)^+,&
 C&=D-H,
 \end{aligned}
\]
where $Y'_e$ is an independent copy of $Y_e$, independent also of
$Y_{-e}$. Thus $D$ is the total deletion surplus, while $H$ counts the
part that survives a second independent draw of the element's own value.
We also define
\[
 x_e(z)=\Prb(e\in B(Y),\ Y_e>z),\qquad
 Q=\sum_e\int_0^\infty x_e(z)^2\,dz.
\]
The quantity $Q$ measures overlap between the value profiles of two
independent optimum bases.

For a height $z>0$, put $p_e(z)=\Prb(Y_e>z)$ and
$S_z=\{e:Y_e>z\}$, and let
\[
 \rho_e(z)=\Prb(e\notin\cl(S_z\setminus e)).
\]
Greedy selects exactly $r(S_z)$ elements above $z$. To identify the
other two integrands, fix $Y_{-e}$. On $A=S_z\setminus e$ the ranks
in $M\setminus e$ and $M/e$ differ by $1$ precisely when $A$ spans
$e$. Integrating this rank difference gives $\tau_e$. Since the sets
$S_z\setminus e$ are nested as $z$ increases, $\tau_e$ is the length
of the initial interval of heights on which they span $e$.
Thus, up to its endpoint, the event $z>\tau_e$ is exactly the event
that $S_z\setminus e$ does not span $e$. Independence now gives
\begin{equation}\label{eq:reference-layer-identities}
 \begin{aligned}
 F&=\int_0^\infty\E r(S_z)\,dz,\\
 D&=\int_0^\infty\sum_e p_e(z)\rho_e(z)\,dz,\\
 H&=\int_0^\infty\sum_e p_e(z)^2\rho_e(z)\,dz.
 \end{aligned}
\end{equation}
The integrand defining $D$ is the expected number of coloops in the
active restriction. Consequently
\begin{equation}\label{eq:structural-basic-bounds}
 0\le H\le D\le F,
 \qquad 0\le Q\le F.
\end{equation}
For the second bound, use $x_e(z)^2\le x_e(z)$ and
$\sum_e x_e(z)=\E r(S_z)$.

Given the actual distributions $X$, the reference distribution with parameter
$q>0$ is the product law $Y^q$ defined by
\begin{equation}\label{eq:reference-tails}
 \Prb(Y_e^q>z)=\min\{1,q\Prb(X_e>z)\},\qquad z\ge0.
\end{equation}
For $q\le1$, this means independently retaining each $X_e$ with probability $q$ and replacing it by zero otherwise. For $q>1$ it puts more
mass on the upper quantiles. Use subscripts $q$ for the quantities of
this reference instance, such as $F_q,D_q,H_q$, and $Q_q$.

The next lemma relates a change in $q$ to deletion surplus. The error
term $H_q$ accounts for coordinates whose reference tail has reached one.

\begin{lemma}[Reference derivatives]\label{lem:reference-derivative}
The function $q\mapsto F_q$ is nondecreasing, continuous, and concave.
At every $q>0$, its finite one-sided derivatives satisfy
\[
 D_q-H_q\le q(F_q)'_+\le q(F_q)'_-\le D_q.
\]
Moreover,
\begin{equation}\label{eq:reference-comparison}
 \min\{1,q\}F_1\le F_q\le\max\{1,q\}F_1.
\end{equation}
\end{lemma}

\begin{proof}
At a fixed height $z$, expected rank is the multilinear extension of
matroid rank evaluated at
$p_e^q(z)=\min\{1,qp_e(z)\}$, where here
$p_e(z)=\Prb(X_e>z)$. Its derivative in coordinate $e$ is the
nonspanning probability $\rho_e^q(z)$. If $qp_e(z)<1$, then
$q\,\partial_q p_e^q(z)=p_e^q(z)$; if $qp_e(z)>1$, this derivative
is zero. At equality, the left derivative includes the coordinate and
the right derivative excludes it. Thus each one-sided derivative,
multiplied by $q$, differs from $D_q$ only by saturated-coordinate
terms. Those terms are bounded by $H_q$, because
$p_e^q(z)^2=p_e^q(z)=1$ on a saturated coordinate.

Both the slope of an unsaturated coordinate and its nonspanning
probability are nonincreasing in $q$. The derivative of each layer's
expected rank is therefore nonincreasing, which proves concavity.
Monotonicity follows by coupling the active sets. Differentiation and
continuity under the height integral are justified locally by the
integrable bound $\sum_e p_e(z)$; every nonloop satisfies
$\E X_e\le F_1$.

For $q\le1$, independently thin an $X$-optimal basis to obtain
$F_q\ge qF_1$, while monotonicity gives $F_q\le F_1$.
Also $F_q\le q\sum_e\E X_e\to0$ as $q\downarrow0$.
Concavity with this zero limit gives $F_q\le qF_1$ for $q\ge1$;
the remaining inequality follows from monotonicity.
\end{proof}

\begin{lemma}[An affine potential]\label{lem:affine-potential}
Let $a$ be a positive nonincreasing affine function on an interval
$J\subset(0,\infty)$. Then $q\mapsto a(q)F_q$ is concave on $J$.
\end{lemma}

\begin{proof}
Let $x<y$ belong to $J$ and put $q=tx+(1-t)y$, with $t\in[0,1]$.
Using concavity of $F$ and then affinity of $a$,
\[
 \begin{aligned}
 a(q)F_q
 &\ge a(q)\bigl(tF_x+(1-t)F_y\bigr)\\
 &=t a(x)F_x+(1-t)a(y)F_y
   +t(1-t)(a(y)-a(x))(F_x-F_y)\\
 &\ge t a(x)F_x+(1-t)a(y)F_y.
 \end{aligned}
\]
The last product is nonnegative because $a$ decreases and $F$ increases.
\end{proof}

\subsection{Stable surplus implies overlap}
\label{sec:stable-overlap}

The derivative bounds involve $H$, but the continuation earns
additional reward in proportion to $Q$. To relate them, run greedy
twice on the same active set, once in decreasing priority order and
once in reverse. An active coloop must survive both scans.

\begin{lemma}[Stable surplus and overlap]\label{lem:stable-overlap}
For every independent nonnegative value vector,
\begin{equation}\label{eq:strong-stable-overlap}
 H^2\le Q(F-D+H)\le FQ.
\end{equation}
\end{lemma}

\begin{proof}
Fix a height $z$ and suppress it from the notation. Let $S$ activate
the labels independently with probabilities $p_e=\Prb(Y_e>z)$.
Independently of these activity indicators, draw for each label a
latent value from its conditional law given $Y_e>z$. Labels with
$p_e=0$ can be omitted. Order the latent values decreasingly, breaking
ties by the fixed label order.

For fixed latent priorities, let $a_e^+$ be the probability that the
active labels preceding $e$ do not span it, and let $a_e^-$ be the
corresponding probability for the labels following it. These events
depend on disjoint sets of independent activity indicators. Nonspan by
all other active labels implies both events, so
\[
 \rho_e\le a_e^+a_e^-.
\]
The function $a_e^+$ is nondecreasing in $e$'s latent value and
nonincreasing in each other latent value; $a_e^-$ has the opposite
monotonicities. Product association, after reversing the order of the
other coordinates, therefore gives
\[
 \E[a_e^+a_e^-]\le(\E a_e^+)(\E a_e^-).
\]
For completeness, association in one coordinate follows from
\[
 2\operatorname{Cov}(f(Z),g(Z))
 =\E[(f(Z)-f(Z'))(g(Z)-g(Z'))]
\]
for an independent copy $Z'$. Conditional covariance and induction
extend it to independent coordinates; replacing one function by its
negative gives the displayed reversed inequality.

Let $x_e^+$ and $x_e^-$ be the inclusion probabilities in the greedy
and reverse-greedy bases of $S$. The reverse order reverses the entire
value-and-label order. Then
\[
 p_e^2\rho_e\le x_e^+x_e^-,\qquad
 \sum_e x_e^+=\sum_e x_e^-=\E r(S).
\]
Here $x_e^+=x_e(z)$ because the greedy choices above height $z$
depend only on the active elements.

Every active coloop belongs to every basis of $S$. Hence
$p_e\rho_e\le x_e^-\le p_e$, and consequently
\[
 x_e^-(1-x_e^-)\ge p_e(1-p_e)\rho_e.
\]
Define
$Q_- =\int_0^\infty\sum_e x_e^-(z)^2\,dz$.
Integrating the preceding inequality and using
\eqref{eq:reference-layer-identities} yields
\[
 F-Q_-\ge D-H.
\]
All these integrals are finite, since the squared marginals are bounded
by the marginals and their sum integrates to $F$. A final
Cauchy--Schwarz inequality gives
\[
 H\le\int_0^\infty\sum_e x_e^+(z)x_e^-(z)\,dz
   \le\sqrt{Q Q_-}\le\sqrt{Q(F-D+H)}.
\]
The second inequality in~\eqref{eq:strong-stable-overlap} follows from
$H\le D$.
\end{proof}

\subsection{Expected contraction prices}

We next relate deletion surplus to the surplus above deterministic
thresholds. The appropriate deterministic prices are expected
contraction costs:
\[
 \pi_e=\E\max\{Y_e,\tau_e\}=m_e+d_e.
\]
Let $B$ be a maximum-$\pi$ basis, and write
\[
 P=\sum_{e\in B}\pi_e,\qquad G=F-P.
\]
The choice of a fixed basis will let us integrate a rank inequality
over all heights. The gap $G$ measures how much we lose by charging
the contraction prices of this basis against the optimum.

\begin{lemma}[Expected contraction prices]\label{lem:core-price-balance}
One has $0\le P\le F$. Every strict superlevel set $T$ of $\pi$
is spanned by $B\cap T$, and $B\cap T$ is a basis of $T$.
Moreover,
\begin{equation}\label{eq:core-coordinate-gap}
 G=\int_0^\infty
   \E\bigl[r(S_z)-|B\cap\cl(S_z)|\bigr]\,dz,
 \qquad S_z=\{e:Y_e>z\},
\end{equation}
and its integrand is nonnegative.
\end{lemma}

\begin{proof}
For deterministic nonnegative weights $y$, the weighted-rank formula
and the rank formula for contraction give
\[
 \OPT(M,y)=\int_0^\infty r(S_z(y))\,dz,
 \quad
 \OPT(M,y)-\OPT(M/e,y)
   =\int_0^\infty\mathbf1\{e\in\cl(S_z(y))\}\,dz.
\]
If $I$ is independent, then $I\cap\cl(S_z(y))$ is independent in
a flat of rank $r(S_z(y))$. Therefore the sum of the realized
contraction costs over $I$ is at most $\OPT(M,y)$.
Take expectations and maximize over $I$ to obtain $P\le F$.
Apply the same identity to the fixed basis $B$ and subtract its price
sum from $F$ to obtain~\eqref{eq:core-coordinate-gap}.
Finally, the greedy characterization of a maximum-weight basis shows
that $B\cap T$ is a basis of every strict superlevel set $T$ of $\pi$.
\end{proof}

Fix $b\in(0,1)$ and cap each reference value at $b\pi_e$:
\[
 \widetilde Y_e=\min\{Y_e,b\pi_e\},\qquad
 \Delta_b=bP-\E\OPT(M,\widetilde Y)\ge0.
\]
The cap deficit is nonnegative because
$\OPT(M,\widetilde Y)\le\OPT(M,b\pi)=bP$.
Define the two threshold surpluses and the remaining capped surplus by
\[
 \begin{aligned}
 S_b&=\sum_e\E(Y_e-bm_e)^+,&
 S_{\pi,b}&=\sum_e\E(Y_e-b\pi_e)^+,\\
 K_b&=\sum_e\E(\min\{Y_e,b\pi_e\}-\tau_e)^+.
 \end{aligned}
\]
In $K_b$, compute $\tau_e$ from the other coordinates of the original
reference vector $Y$.

\begin{lemma}[Surplus decomposition]\label{lem:core-pricing}
\begin{equation}\label{eq:core-pricing-bound}
 S_b\ge S_{\pi,b}\ge(1-b)F+bG+\Delta_b,
 \qquad D\le S_{\pi,b}+K_b.
\end{equation}
\end{lemma}

\begin{proof}
The first comparison follows from $m_e\le\pi_e$. Evaluating the capped
weights on a $y$-optimal basis gives, for every realization,
\[
 \OPT(M,y)\le\OPT(M,\min\{y,b\pi\})
                   +\sum_e(y_e-b\pi_e)^+.
\]
Take expectations and use $P=F-G$ for the second comparison.
For the last inequality, apply
$(y-t)^+\le(y-w)^++(\min\{y,w\}-t)^+$
with $t=\tau_e$ and $w=b\pi_e$, and sum over $e$.
\end{proof}

\subsection{A rank inequality from Poisson concavity}

To bound $K_b$, we use the concavity of Poisson-rounded matroid rank,
which underlies the rounding method of Dughmi, Roughgarden, and
Yan~\cite{DRY2011}. A tangent inequality evaluated at one scaled basis vector
suffices here. The parameter $s$ below will be chosen in Appendix~\ref{sec:parameters}.

\begin{lemma}[A fixed Poisson tangent]\label{lem:core-poisson-tangent}
Let $N_0$ be a rank-$r_0$ matroid with a basis $C_0$, and let $R$
activate labels independently with probabilities $p_e$. Put
\[
 \rho_e=\Prb(e\notin\cl_{N_0}(R\setminus e)),\qquad
 m=\sum_{f\in C_0}\Prb(f\notin\cl_{N_0}(R)).
\]
For every $s>0$,
\begin{equation}\label{eq:fixed-poisson-tangent}
 \sum_e p_e\rho_e
 \le\sum_e p_e^2\rho_e+r_0e^{-s}+s m.
\end{equation}
\end{lemma}

\begin{proof}
First suppose $p_e<1$ and set $\lambda_e=-\log(1-p_e)$.
For $x\in\mathbb R_{\ge0}^{E(N_0)}$, let $R_x$ activate each label
independently with probability $1-e^{-x_e}$, and define
$f(x)=\E r_{N_0}(R_x)$. Direct differentiation gives
\[
 \partial_e f(x)=\Prb(e\notin\cl_{N_0}(R_x)).
\]
The diagonal second derivative is the negative of this probability.
For distinct labels $e,f$, the mixed second derivative is minus the
probability that $e$ and $f$ are parallel nonloops in $N_0/R_x$,
with contracted labels treated as loops. To see this, condition on
the other coordinates. The mixed discrete rank difference is $-1$
exactly for a parallel pair of nonloops; the factors
$e^{-x_e-x_f}$ enforce the absence of $e,f$ from $R_x$.

Thus the negative Hessian is the expectation of a block matrix with one
all-ones block for each nonloop parallel class of $N_0/R_x$ and zeros
elsewhere. Each block is positive semidefinite, so $f$ is concave.
At $x=\lambda$, write
$u_e=\partial_e f(\lambda)=(1-p_e)\rho_e$.
Since $C_0$ is a basis,
$f(s\mathbf1_{C_0})=r_0(1-e^{-s})$. The tangent inequality gives
\[
 \sum_e\lambda_e u_e
 \le f(\lambda)-r_0(1-e^{-s})+s\sum_{f\in C_0}u_f
 \le r_0e^{-s}+s m.
\]
Finally, $p_e=1-e^{-\lambda_e}\le\lambda_e$, and hence
\[
 \sum_e p_e\rho_e-\sum_e p_e^2\rho_e
 =\sum_e p_eu_e\le\sum_e\lambda_eu_e.
\]
This proves the claim. Coordinates with $p_e=1$ follow by continuity
of the finite Bernoulli expectations. Rank zero is immediate.
\end{proof}

\subsection{Bounding the capped surplus}

We apply the rank inequality separately at every value layer. At height $z$ the caps retain only those labels with $b\pi_e>z$,
but an uncapped active label outside this set can still span a label
inside. Conditioning on the outside active labels and contracting them
keeps this dependence in the rank calculation.

\begin{lemma}[Capped surplus]\label{lem:core-fixed-tangent}
For every $b\in(0,1)$ and $s>0$,
\begin{equation}\label{eq:core-fixed-tangent-main}
 K_b\le H+bP e^{-s}+s(G+\Delta_b).
\end{equation}
\end{lemma}

\begin{proof}
At height $z>0$, let $T_z=\{e:b\pi_e>z\}$. Independence of
$Y_e$ from $\tau_e$ gives
\[
 K_b=\int_0^\infty\sum_{e\in T_z}p_e(z)\rho_e(z)\,dz.
\]
By Lemma~\ref{lem:core-price-balance}, $B\cap T_z$ is a basis of
$T_z$. Condition on the outside active set $O=S_z\setminus T_z$.
The indicators inside $T_z$ remain independent. The minor
$(M/O)|_{T_z}$ has rank at most $r(T_z)$. Since $B\cap T_z$ spans
$T_z$ in $M$, it still spans $T_z$ in this minor. Choose any basis
of the minor contained in $B\cap T_z$. For this choice the quantity
$m$ in Lemma~\ref{lem:core-poisson-tangent} is at most
\[
 m_z(O)=\sum_{f\in B\cap T_z}
                    \Prb(f\notin\cl(S_z)\mid O).
\]
Apply Lemma~\ref{lem:core-poisson-tangent} to this minor and average
over $O$. The two error terms are linear, so
\[
 \sum_{e\in T_z}p_e(z)\rho_e(z)
 \le\sum_{e\in T_z}p_e(z)^2\rho_e(z)
        +e^{-s}r(T_z)+s m(z),
\]
where
$m(z)=\sum_{f\in B\cap T_z}\Prb(f\notin\cl(S_z))$.

It remains to bound the integrated missing mass. For every
deterministic set $S$, the set
$(B\setminus T_z)\cap\cl(S)$ is independent over $T_z$:
$B\cap T_z$ spans $T_z$, and $B$ is independent. It is therefore
also independent over $S\cap T_z$. As all its elements belong to
$\cl(S)$,
\[
 r(S)\ge r(S\cap T_z)+|(B\setminus T_z)\cap\cl(S)|.
\]
Equivalently,
\begin{equation}\label{eq:core-missing-mass-pointwise}
 r(T_z)-|B\cap T_z\cap\cl(S)|
 \le r(T_z)-r(S\cap T_z)+r(S)-|B\cap\cl(S)|.
\end{equation}
Average and integrate. The left side integrates to
$\int_0^\infty m(z)\,dz$. The first rank deficit integrates to
$\Delta_b$, because the capped active set is $S_z\cap T_z$ and
$\int_0^\infty r(T_z)\,dz=bP$. The second integrates to $G$ by
\eqref{eq:core-coordinate-gap}. Therefore
\[
 \int_0^\infty m(z)\,dz\le G+\Delta_b.
\]
Now integrate the conditional rank inequality and enlarge its
restricted squared-probability term to $H$.
\end{proof}

The two deficits $G$ and $\Delta_b$ can now be eliminated. The resulting
bound depends only on the quantities used to select the reference
distribution and the thresholds used by the online policy.

\begin{corollary}[Threshold surplus]\label{cor:structural-surplus}
If $F>0$, then for every $b\in(0,1)$ and $s>0$,
\begin{equation}\label{eq:structural-surplus}
 \frac{S_b}{F}\ge1-b+
 \frac{b}{b+s}
 \left(\frac DF-(1-b)-\frac HF-b e^{-s}\right).
\end{equation}
\end{corollary}

\begin{proof}
Lemma~\ref{lem:core-pricing} implies
\[
 G+\Delta_b\le\frac{S_b-(1-b)F}{b}.
\]
Combining the same lemma with Lemma~\ref{lem:core-fixed-tangent} and
$P\le F$ gives
\[
 D\le S_b+H+bF e^{-s}
             +\frac{s}{b}\bigl(S_b-(1-b)F\bigr).
\]
Rearrange and divide by $F$.
\end{proof}

All layer integrals above allow atoms. For each realization, changing
a strict height cutoff to a weak one affects only a finite set of
heights. Tonelli's theorem therefore gives the same integrals after
taking expectations. Integrability follows from $F<\infty$, since the
mean of each nonloop is at most $F$.

\section{A virtual basis-exchange continuation}\label{sec:controller}

Suppose two independent reference optima often contain the same
valuable elements. We will use this overlap late in the arrival
sequence. Each accepted label is inserted into every reference basis
by an exchange. The algorithm stores only one record per acceptance.
It can then recover any particular basis by replaying the exchanges
starting from its original greedy basis.

As in the survival simulation of~\cite[Section~6]{ABHHMS2026}, the
algorithm tests a survival event on an auxiliary random sample.
Here a test draws a reference basis and replays the accepted records.
Lemma~\ref{lem:hazard-identity} gives the resulting acceptance rates.

Fix a loopless residual matroid $M$ on $n$ labels, a parameter $q\in J$,
and the product reference law $\mu$ of $Y=Y^q$. We first describe
the algorithm for finitely supported distributions, so that its
screening probabilities are ratios of point masses. The approximation
argument at the end of Appendix~\ref{sec:parameters} extends the
existence guarantee to general distributions. Write $D_e$ and
$D_e^q$ for the original and reference distributions. Put
\begin{equation}\label{eq:controller-parameters}
 \lambda=\frac1q,\qquad t_0=1-\delta,\qquad
 \alpha=e^{-\lambda t_0},\qquad \omega=\alpha-\lambda\delta,
\end{equation}
and assume $\omega>0$. Define
\begin{equation}\label{eq:controller-phi}
 \Phi_\delta(q,z)
 =1-\alpha+\alpha\omega(1-e^{-\lambda\delta/\omega})
                         \bigl(1+(1-\alpha)z\bigr).
\end{equation}
For $\delta=0$, this is the familiar $1-e^{-1/q}$ guarantee.
For a short positive final interval, its dependence on $z$ will give
the improvement.

\begin{proposition}[Virtual continuation]\label{prop:controller}
If $F=\E\OPT(M,Y)>0$, there is a feasible online continuation with
\[
                   \E[\ALG]\ge\Phi_\delta(q,Q/F)F.
\]
In the exact-randomness model it uses $O(n^4)$ matroid independence queries
in expectation. Its decisions use only the current arrival and
previously accepted records.
\end{proposition}

\subsection{Updating virtual bases}

Fix the label order from Section~\ref{sec:preliminaries}.
For each pair of bases $C,B$, we need one injective exchange map
from $C$ to $B$. The same map must be used regardless of which
source element is queried.

\begin{lemma}[A canonical exchange map]\label{lem:exchange-map}
For bases $C,B$, one can construct a bijection
$\phi_{C,B}:C\to B$ fixing $C\cap B$ pointwise such that
\[
                   B-\phi_{C,B}(f)+f\in\mathcal B(M)
                   \qquad(f\in C).
\]
The map uses $O(n^2)$ matroid independence queries.
\end{lemma}

\begin{proof}
Keep the target basis $B$ fixed and initialize a working basis
$C'=C$. Fix every element of $C\cap B$. Until $C'=B$, take the first
$f\in C'\setminus B$ in the label order and scan $b\in B\setminus C'$
in that order. Choose the first $b$ such that both $B-b+f$ and
$C'-f+b$ are bases. Assign $\phi_{C,B}(f)=b$ and replace $C'$ by
$C'-f+b$.

Such a $b$ exists. Let $\Gamma\subseteq B+f$ be the fundamental
circuit of $f$, and let $H=\cl(C'\setminus f)$. Since $f\notin H$
and $f\in\cl(\Gamma\setminus f)$, some $b\in\Gamma\setminus f$
lies outside $H$. This element belongs to $B\setminus C'$ and
permits both exchanges.

Each step removes one original source element outside $B$ and adds
one previously absent element of $B$. Thus the assigned images do
not repeat, and the procedure ends at $B$. Every image satisfies
the required exchange property with respect to the unchanged
target. If $d=|C\setminus B|$, at most
$2(d+(d-1)+\cdots+1)=d(d+1)$ queries are used.
\end{proof}

To evaluate only $\phi_{C,B}(e)$, the algorithm stops this fixed scan
when $e$'s image is assigned. The scan order does not depend on $e$.
This consistency is what allows us to charge all possible insertions
to distinct target elements.

\paragraph{Reference realizations and accepted records.}
For a reference realization $y$, let $B_y^0$ be its canonical
maximum-weight basis. The continuation algorithm stores records
$(e_i,t_i,C_i)$, one for each acceptance, where $C_i$ is the
candidate basis that justified that acceptance.
Let $\mathcal H$ be the ordered list of accepted records. Define
$\operatorname{Replay}(y,\mathcal H,t)$ by starting with $B=B_y^0$
and replaying the accepted records with $t_i\le t$ in order:
\[
                         B\leftarrow B-\phi_{C_i,B}(e_i)+e_i.
\]
Write $B_y(t)=\operatorname{Replay}(y,\mathcal H,t)$. At an
arrival, evaluate this basis using the records of earlier acceptances.
The continuation always samples from the same law on the same
matroid $M$. In particular, a fresh reference basis may contain
a previously rejected label.

Every current basis contains the accepted set $A_t$. Indeed, the
source and target bases at an insertion both contain the previous
accepted set, which the exchange map fixes pointwise. A pair $(e,y)$
is \emph{live} at $t$ if $y_e>0$ and $e\in B_y(t)\setminus A_t$.
Only accepted labels are inserted, so an unaccepted label present
in $B_y(t)$ also belongs to $B_y^0$. Once a live pair disappears
it cannot become live again.

For a fixed accepted history, define
\begin{equation}\label{eq:live-kernels}
 \begin{aligned}
 \mathcal K_{e,t}(dy)&=\one\{(e,y)\text{ is live at }t\}\mu(dy),\\
 k_e(t)&=\int\mathcal K_{e,t}(dy),&
 w_e(t)&=\int y_e\,\mathcal K_{e,t}(dy).
 \end{aligned}
\end{equation}
These are analysis quantities. The algorithm tests liveness in
sampled reference realizations and never evaluates these integrals.
For every reference realization, the live labels form a subset of $B_y^0$ and of
$B_y(t)\setminus A_t$. Consequently they are feasible in $M/A_t$,
and $\sum_e w_e(t)\le F$.

\subsection{The arrival rule}

Assign independent uniform times in $[0,1]$ to the residual labels.
In the random-permutation model, this means drawing and sorting $n$
independent uniform numbers, then attaching them in order to the
remaining arrivals. The rule uses the current time and accepted
records; unused future clocks and rejected observations do not
enter its decisions.

\begin{algorithm}[t]
\caption{Virtual continuation: processing an arrival $(e,v,t)$}
\label{alg:controller}
\textbf{Fixed input:} residual matroid $M$, parameter $q$, distributions
$D_f,D_f^q$, and switching time $t_0=1-\delta$.\\
\textbf{State:} accepted set $A$ and ordered record list $\mathcal H$,
both initially empty on entering the continuation.\\
\textbf{Output:} accept or reject $e$, with an updated state on acceptance.
\begin{enumerate}[leftmargin=*,itemsep=3pt,topsep=5pt]
\item \textbf{Screen the value.} If $v=0$, \textbf{return reject}.
Draw $B_{\rm scr}\sim\operatorname{Bernoulli}
\bigl(D_e^q(\{v\})/(qD_e(\{v\}))\bigr)$.
If $B_{\rm scr}=0$, \textbf{return reject}.
\item \textbf{Sample a candidate basis.} Set $y_e=v$ and draw
$y_f\sim D_f^q$ independently for all $f\ne e$.
Set $C\gets\operatorname{Replay}(y,\mathcal H,t)$.
If $e\notin C\setminus A$, \textbf{return reject}.
\item \textbf{Test early acceptance.} If $t\le t_0$:
\begin{enumerate}[label=(\alph*),leftmargin=1.6em,itemsep=2pt,topsep=2pt]
\item Draw $Z\sim\operatorname{Poisson}(2)$.
\item For $j=1,\ldots,Z$:
\begin{enumerate}[label=(\roman*),leftmargin=1.8em,itemsep=2pt,topsep=2pt]
\item Draw $U_j$ uniformly from $[0,1)$ and set $u_j\gets2qU_j$.
\item If $u_j<t$, draw a fresh full reference vector
$z\sim\prod_fD_f^q$, set
$C'\gets\operatorname{Replay}(z,\mathcal H,u_j)$, and let
$A(u_j)=\{e_i:(e_i,t_i,C_i)\in\mathcal H,\ t_i\le u_j\}$.
If $z_e>0$ and $e\in C'\setminus A(u_j)$,
\textbf{return reject}.
\end{enumerate}
\end{enumerate}
\item \textbf{Accept and update.} Set $A\gets A\cup\{e\}$,
append $(e,t,C)$ to $\mathcal H$, earn reward $v$, and
\textbf{return accept}.
\end{enumerate}
\end{algorithm}

Candidate and probe draws are independent. A candidate uses the
observed $e$-coordinate, whereas each probe samples that coordinate
afresh. After $t_0$, the algorithm accepts every screened live
candidate.

The reference screen is a valid probability. The function
$u\mapsto\min\{1,qu\}$ is $q$-Lipschitz, so each positive reference
atom has mass at most $q$ times the corresponding original atom.
Multiplying the original atom by the screen probability and sampling
the other coordinates gives the candidate-realization submeasure
$\lambda\mathcal K_{e,t}$ before the Poisson test.

Every acceptance is feasible because its candidate basis contains
all earlier accepted labels. Evaluating a fresh reference realization requires a greedy scan
and at most $n$ exchange replays, hence $O(n^3)$ queries.
There is at most one candidate realization and an expected two probe
realizations per arrival, giving $O(n^4)$ expected queries. It remains to
prove the reward guarantee.

\subsection{The law of the accepted process}

A missing label may still be on its way, or it may already have
been rejected. We keep both possibilities when conditioning on
accepted records. To state the calculation, attach to each acceptance
a \emph{mark}, such as the value and the reference realization used
in the decision. The history records the labels, times, and marks of
acceptances. Given this history, the \emph{intensity} of an acceptance
with mark in a set is its conditional probability per unit time.

\begin{lemma}[Acceptance intensity]\label{lem:hazard-identity}
For each label $e$, let $T_e$ be a uniform arrival time, $X_e$ its
actual value, and $R_e$ a private random tape. These triples are
independent across labels, and their coordinates are independent.
Suppose the decision at an arrival uses only the accepted marked
history. For a frozen prefix $h$, let
$\Psi_e(h,t,X_e,R_e)$ be either rejection or an acceptance mark, and
let $\nu_{e,h,t}$ be its acceptance subprobability kernel under the
unconditional law of $(X_e,R_e)$.

Along a prescribed accepted path, write $h_t$ for the prefix before
$t$ and, for a label not yet accepted, put
\[
             S_e(t)=1-\int_0^t\nu_{e,h_u,u}(\mathcal M)\,du,
\]
where $\mathcal M$ is the acceptance-mark space. Wherever $S_e>0$,
the marked acceptance intensity, with respect to the history of
accepted marks, is $\nu_{e,h_t,t}/S_e(t)$.
\end{lemma}

\begin{proof}
Fix a marked accepted path through time $t$. A label absent from
this path satisfies precisely
\[
 \{T_e>t\}\ \cup\
 \{T_e\le t,\ \Psi_e(h_{T_e},T_e,X_e,R_e)
                                     \text{ is rejection}\}.
\]
Its probability is $S_e(t)$. The sole arrival time is integrated
once; no independence of decisions at counterfactual times is
asserted. A label accepted at time $t_i$ with mark $m_i$ contributes
the event density $\nu_{e_i,h_{t_i},t_i}(dm_i)\,dt_i$.
Once the path is fixed, these are constraints on independent
per-label triples. Scanning actual arrivals in order shows that
the constraints are also sufficient to realize that path.

In the process with the claimed intensities,
$-d\log S_e(u)=\nu_{e,h_u,u}(\mathcal M)\,du/S_e(u)$ before $e$
is accepted. The survival factor is $S_e(t)$ for an absent label.
For an accepted label it is $S_e(t_i)$, which cancels the
denominator in its event intensity. Thus the two processes have
the same marked-path densities. There are at most $n$ events, so
these densities determine their laws.
\end{proof}

For analysis, include the credited value and the complete candidate
realization in each acceptance mark. The other coordinates of that realization
come from the accepting label's private tape; they are not other
labels' actual values. Thus the preceding factorization still
applies. Write $\mathcal F_t$ for the filtration generated by the
accepted marks, that is, the information they reveal through time $t$.
The implemented state only needs the recorded bases. All kernels
below use the prefix before a decision, or equivalently their left
limits at an event time.

Set
\[
                         \Lambda_e(t)=\lambda\int_0^t k_e(u)\,du.
\]
Since $2q\ge1$, a raw early probe finds $e$ live with probability
$\Lambda_e(t)/2$. The Poisson test passes with probability
$e^{-\Lambda_e(t)}$. Hence the actual acceptance kernel on candidate realizations is
$\lambda e^{-\Lambda_e(t)}\mathcal K_{e,t}$ in the early phase,
and $\lambda\mathcal K_{e,t}$ in the late phase.
Mapping each candidate realization to its acceptance mark gives
the kernel in Lemma~\ref{lem:hazard-identity}. For labels not yet accepted, that
lemma gives
\[
 S_e(t)=
 \begin{cases}
   e^{-\Lambda_e(t)},&t\le t_0,\\[1mm]
   e^{-\Lambda_e(t_0)}-\lambda\displaystyle\int_{t_0}^t k_e(u)\,du,
      &t>t_0.
 \end{cases}
\]
Thus the early marked acceptance intensity is exactly
$\lambda\mathcal K_{e,t}$. In the late phase it is
$r_e(t)\mathcal K_{e,t}$, where
\begin{equation}\label{eq:late-rate-bounds}
 r_e(t)=\frac{\lambda}{S_e(t)},\qquad
 \lambda(1+\Lambda_e(t_0))\le r_e(t)\le\frac{\lambda}{\omega}.
\end{equation}
The lower bound uses $e^x\ge1+x$. For the upper bound, use
$k_e\le1$ and $\Lambda_e(t_0)\le\lambda t_0$ to obtain
$S_e(t)\ge\alpha-\lambda\delta=\omega>0$.
After a label is accepted its kernels and rates are set to zero.
These rates are only for the proof: the late rule is simply to
accept every screened live candidate.

\subsection{Persistence and the overlap gain}

\begin{lemma}[Persistence]\label{lem:persistence}
If the marked acceptance intensities are $r_e(t)\mathcal K_{e,t}$ and
$r_e(t)\le R$ throughout an interval, every live pair has conditional
death rate at most $R$, where the death rate is the instantaneous
conditional rate at which the pair ceases to be live. In particular, for $u\le v$ in that interval,
\[
                \E[w_e(v)\mid\mathcal F_u]\ge e^{-R(v-u)}w_e(u).
\]
The same inequality holds for the mass of live pairs with $y_e>z$.
\end{lemma}

\begin{proof}
Fix a live target pair and its current basis $B$. In any source
realization with current basis $C$, injectivity of $\phi_{C,B}$ allows
at most one live source label to remove the target element.
If the target label itself is live in the source, the map fixes it
and no other source label can remove it. Its own acceptance is
included in this bound. The killing intensity from that source
realization is therefore at most $R\mu(dy)$. Integration over source
realizations gives the death-rate bound. Now fix a target realization
live at $u$. Its conditional survival
probability $p(v)$ has $p(u)=1$ and $p'(v)\ge -Rp(v)$ almost
everywhere, since disappearance is irreversible. Thus
$p(v)\ge e^{-R(v-u)}$. Integrate over live targets at $u$ with
weight $y_e$ or $\one\{y_e>z\}$ to obtain the claim.
\end{proof}

This proof explains why the early random rejections act on whole
source realizations. A bound on the total acceptance rate of each
label alone would allow several labels to concentrate their mass on
bases whose exchanges all evict the same target.

In the early phase the instantaneous reward rate is
$\lambda\sum_e w_e(t)$. Persistence with $R=\lambda$ and
$\sum_e w_e(0)=F$ bounds its expectation below by
$\lambda e^{-\lambda t}F$. Integration gives expected early reward
at least $(1-\alpha)F$, and at the switching time
\begin{equation}\label{eq:early-mass}
                         \E\sum_e w_e(t_0)\ge\alpha F.
\end{equation}
We also claim
\begin{equation}\label{eq:overlap-accumulation}
 \E\sum_e w_e(t_0)\Lambda_e(t_0)\ge\alpha(1-\alpha)Q.
\end{equation}
To prove this, let
$u_e(t,z)=\int\one\{y_e>z\}\mathcal K_{e,t}(dy)$.
Since $u_e(t,z)\le k_e(t)$,
\[
              w_e(t)k_e(t)\ge\int_0^\infty u_e(t,z)^2\,dz.
\]
Persistence gives $\E u_e(t,z)\ge e^{-\lambda t}x_e(z)$.
Conditional persistence followed by Jensen's inequality therefore
yields
\begin{align*}
 \E\sum_e w_e(t_0)\Lambda_e(t_0)
 &\ge\lambda\int_0^{t_0}e^{-\lambda(t_0-t)}
                         \E\sum_e w_e(t)k_e(t)\,dt\\
 &\ge\lambda\int_0^{t_0}e^{-\lambda(t_0-t)}e^{-2\lambda t}Q\,dt
   =\alpha(1-\alpha)Q.
\end{align*}

\begin{proof}[Proof of Proposition~\ref{prop:controller}]
By~\eqref{eq:late-rate-bounds}, late persistence holds with
$R=\lambda/\omega$, while the credit intensity is at least
$\lambda\sum_e(1+\Lambda_e(t_0))w_e(t)$.
Conditional on $\mathcal F_{t_0}$, the late reward is consequently
at least
\[
        \omega(1-e^{-\lambda\delta/\omega})
                        \sum_e(1+\Lambda_e(t_0))w_e(t_0).
\]
Add the early reward and use~\eqref{eq:early-mass}
and~\eqref{eq:overlap-accumulation}. The result is
\[
 \E[\ALG]\ge
 (1-\alpha)F+\alpha\omega(1-e^{-\lambda\delta/\omega})
                                  \bigl(F+(1-\alpha)Q\bigr),
\]
which is exactly~\eqref{eq:controller-phi}. Feasibility and the
query bound were established above.
\end{proof}

\section{Estimating a reference instance}
\label{sec:combined-estimation}

All quantities used by the search have a rank-layer representation.
This permits a common estimator even when a large value occurs with
very small probability. Sample the height of a layer in proportion to
the probability that it contains an active element, then sample that
layer conditioned on being nonempty. The observations are bounded in
units of the expected maximum, regardless of the largest support value.

Throughout the implementation let $N$ denote the original ground-set
size. Fix a residual matroid on $n\le N$ nonloops and a reference
$q\in J$. Write
\[
 V_q=\E\max_e Y_e^q,\qquad C_q=D_q-H_q.
\]
We stop at states with zero benchmark, so $V_q>0$ below.

\subsection{Sampling a nonempty layer}

Sort the distinct positive values appearing in the residual support
lists as $0=v_0<v_1<\cdots<v_s$. Write
$\Delta_j=v_j-v_{j-1}$ and let $p_{e,j}$ be the reference tail
probability on the interval $(v_{j-1},v_j)$. In a fixed label order,
form the joint weights
\[
 w_{j,b}=\Delta_jp_{b,j}\prod_{i<b}(1-p_{i,j}).
\]
Their sum is
\[
 \sum_{j,b}w_{j,b}
 =\sum_j\Delta_j\left(1-\prod_e(1-p_{e,j})\right)=V_q.
\]
Draw $(j,b)$ with probability $w_{j,b}/V_q$. Set the indicators before
$b$ to zero and the indicator of $b$ to one; draw each later indicator $e$ independently
with probability $p_{e,j}$. Denote the resulting
nonempty active set by $R$. If $R_j$ is an unconditioned product layer,
then
\begin{equation}\label{eq:combined-layer-law}
 \Pr((j,R))=\frac{\Delta_j}{V_q}\Pr(R_j=R),\qquad R\ne\varnothing.
\end{equation}
Indeed, $R$ has a unique first active label. Even when $\Pr(R_j\ne\varnothing)$
is small, this construction uses one categorical draw and at most $n$
Bernoulli draws.

Let $\mathcal C(R)$ be the coloops of $M|_R$. From this one layer form
\[
 \begin{aligned}
 Z_F&=V_q r(R), & Z_D&=V_q|\mathcal C(R)|,\\
 Z_C&=V_q\sum_{e\in\mathcal C(R)}(1-p_{e,j}),
 &Z_e&=V_q\mathbf1\{e\in\operatorname{cl}(R\setminus e)\}.
 \end{aligned}
\]

\begin{lemma}[Common layer estimates]\label{lem:combined-layer-estimates}
The expectations of $Z_F,Z_D,Z_C,Z_e$ are $F_q,D_q,C_q,m_e^q$,
respectively. On every outcome,
\[
 V_q\le Z_F\le NV_q,\qquad
 0\le Z_C\le Z_D\le Z_F,\qquad 0\le Z_e\le V_q.
\]
All observations can be computed with $O(n\log(n+1))$ independence
queries.
\end{lemma}

\begin{proof}
The rank layer formula gives the first expectation. An active element
is a coloop of the active restriction exactly when the other active
elements do not span it. Thus the unconditioned expectations of the
next two layer statistics are $\sum_ep_e\rho_e$ and
$\sum_ep_e(1-p_e)\rho_e$, where
$\rho_e=\Pr(e\notin\operatorname{cl}(R_j\setminus e))$.
Their height integrals are $D_q$ and $D_q-H_q$. Similarly,
\[
 m_e^q=\int_0^\infty
     \Pr(e\in\operatorname{cl}(R_z\setminus e))\,dz.
\]
All integrands vanish on the empty layer because the residual matroid
has no loops. Equation~\eqref{eq:combined-layer-law} therefore proves
the asserted expectations. The bounds follow from
$1\le r(R)\le n$ and $|\mathcal C(R)|\le r(R)$.

For the query bound, greedily find a basis $B$ of $R$ and initialize
$U=B$. For $f\in R\setminus B$, its fundamental circuit $C_f$ satisfies
\[
 (B\setminus S)+f\text{ is independent}
 \quad\Longleftrightarrow\quad C_f\cap S\ne\varnothing
 \qquad(S\subseteq B).
\]
Binary subdivision finds a member of $C_f\cap U$, when one exists,
using $O(\log(n+1))$ queries. Remove each such member from $U$ and
continue until the intersection is empty. Every basis element is
removed at most once, and each nonbasis element contributes one final
unsuccessful test. At termination $U=\mathcal C(R)$. For $e\in R$,
the closure indicator is $\mathbf1\{e\notin U\}$; for $e\notin R$,
one query on $B+e$ supplies it. This proves the bound.
\end{proof}

\subsection{A single accuracy guarantee}

Let $K$ be the search budget in~\eqref{eq:combined-search-budget}. Set
\[
 H_0=\left\lceil\log_2\frac{2NK(N+3)}\rho\right\rceil,
 \qquad
 m=\left\lceil\frac{2NH_0}{\varepsilon^2}\right\rceil.
\]
At each evaluated parameter draw exactly $m$ fresh layers and average
the observations. Denote the resulting estimates by
$\widehat F,\widehat D,\widehat C,\widehat m_e$.

\begin{lemma}[Simultaneous accuracy]\label{lem:combined-confidence}
Conditionally at any residual state, with probability at least
$1-\rho/N$, every one of its at most $K$ adaptively requested batches
satisfies
\[
 \begin{aligned}
 |\widehat F-F_q|&\le\varepsilon(F_q+V_q),\\
 |\widehat D-D_q|&\le\varepsilon(D_q+V_q),\\
 |\widehat C-C_q|&\le\varepsilon(C_q+V_q),\\
 |\widehat m_e-m_e^q|&\le\varepsilon m_e^q+\varepsilon V_q/N
                      \qquad(e\in E).
 \end{aligned}
\]
The same bounds hold for the saved batch at an adaptively selected
evaluated parameter.
\end{lemma}

\begin{proof}
For independent $W_i\in[0,1]$ of mean $\mu$, Bernstein's inequality
and $\operatorname{Var}(W_i)\le\mu$ imply
\[
 \Pr\bigl(|\overline W-\mu|>\gamma\mu+t\bigr)
 \le2\exp\left(-\frac{m(\gamma\mu+t)^2}
                         {2\mu+(2/3)(\gamma\mu+t)}\right)
 \le2e^{-m\gamma t/2},\quad 0<\gamma\le1.
\]
The last inequality follows by expanding the two sides of
$(\gamma\mu+t)^2\ge(\gamma t/2)[2\mu+(2/3)(\gamma\mu+t)]$.
Normalize the first three observations by $NV_q$ and each $Z_e$ by
$V_q$, and take $\gamma=\varepsilon$, $t=\varepsilon/N$.
Each statistic fails with probability at most
$\rho/[NK(N+3)]$. A conditional union bound over the at most $N+3$
statistics and $K$ parameter evaluations proves the claim. Indeed, conditional on the
previous evaluations, the next parameter is already fixed. Taking the
intersection of the accuracy events also covers the batch eventually
returned by the search.
\end{proof}

On the accurate event, the upper estimates
\begin{equation}\label{eq:combined-upper-price}
 \overline m_e=
       \frac{\widehat m_e+\varepsilon V_q/N}{1-\varepsilon}
\end{equation}
satisfy
\[
 m_e^q\le\overline m_e
 \le\frac{1+\varepsilon}{1-\varepsilon}m_e^q
       +\frac{2\varepsilon V_q}{N(1-\varepsilon)}.
\]
For the threshold $p_e=a(q)\overline m_e$, put
$b=a(q)(1+\varepsilon)/(1-\varepsilon)$. Our parameter ranges give
$b<1$, $0\le b-a(q)<2\varepsilon$, and
\begin{equation}\label{eq:combined-price-sum}
 p_e\ge a(q)m_e^q,
 \qquad \sum_e(p_e-bm_e^q)^+<2\varepsilon F_q.
\end{equation}
Thus the same batch supplies the search statistics, the stability test,
and all possible next-arrival thresholds.

\section{Choosing the reference parameter}
\label{sec:combined-search}

A small additive error in the reference potential could accumulate over
many arrivals. We avoid this loss by moving the reference only when the
samples certify the sign of a one-sided derivative. At the returned
point we need just two inequalities relating the deletion surplus to
the stable surplus.

Write
\[
 g(q)=a(q)F_q,\qquad
 d_*(q)=\frac{\theta q}{1-\theta q}.
\]
Lemma~\ref{lem:affine-potential} gives concavity of $g$.
Since $a'=-\theta$, Lemma~\ref{lem:reference-derivative} gives
\begin{equation}\label{eq:combined-slope-bracket}
 \frac{aF_q}{q}\left(\frac{C_q}{F_q}-d_*(q)\right)
 \le g'_+(q)\le g'_-(q)
 \le\frac{aF_q}{q}\left(\frac{D_q}{F_q}-d_*(q)\right).
\end{equation}
At an evaluated parameter call the batch positive if
$\widehat C/\widehat F>d_*+10\varepsilon$, and negative if
$\widehat D/\widehat F<d_*-10\varepsilon$. These tests are disjoint,
since $0\le\widehat C\le\widehat D\le\widehat F$ on every outcome.
The denominator is positive because $\widehat F\ge V_q$.

Initialize the search at the inherited $q^{\rm in}$, where
$|q^{\rm in}-1|<3\delta$. Evaluate this parameter. If neither test holds,
return it and its batch. If its batch is positive, bisect
$[q^{\rm in},1+4\delta]$: a positive midpoint replaces the left
endpoint and its saved batch, and any other midpoint replaces the
right endpoint. Return the saved left endpoint. If its batch is negative,
use the symmetric procedure on $[1-4\delta,q^{\rm in}]$ and return
the saved right endpoint. In either branch perform $K-1$ midpoint evaluations, where
\begin{equation}\label{eq:combined-search-budget}
 K=1+\left\lceil
       \log_2\max\left\{1,\frac{32\delta N^2}{\varepsilon}\right\}
       \right\rceil.
\end{equation}
The terminal interval has length at most $\varepsilon/(4N^2)<\delta$.

\begin{lemma}[Reference selection]\label{lem:combined-reference-selection}
On the simultaneous accurate event, the returned reference satisfies
\[
 a(q)F_q\ge a(q^{\rm in})F_{q^{\rm in}}.
\]
If $|q-1|<3\delta$, then
\begin{equation}\label{eq:combined-stationarity}
 d_*(q)-20\varepsilon\le D_q/F_q
 \le d_*(q)+H_q/F_q+20\varepsilon.
\end{equation}
Its saved batch satisfies all bounds of
Lemma~\ref{lem:combined-confidence}.
\end{lemma}

\begin{proof}
Accuracy and $C_q,D_q\le F_q$ imply that both estimated ratios differ
from their true ratios by at most
$4\varepsilon/(1-2\varepsilon)\le5\varepsilon$.
Thus a positive batch certifies $g'_+>0$, and a negative batch certifies $g'_-<0$.
By concavity, a parameter with a positive batch has larger potential than every
point to its left, and a parameter with a negative batch has larger potential
than every point to its right. Every update to the saved endpoint
therefore increases the potential.

The two endpoints of the final interval may have different test
outcomes. To transfer a failed test across that interval, we bound how
fast its true ratio can change. Put $q_-=1-4\delta\ge1/2$ and $a_-=\min_Ja\ge1/2$.
Coupling the product layers at two parameters gives
\[
 |D_q-D_s|\le n^2F_1|q-s|,
 \quad |H_q-H_s|\le2n^2F_1|q-s|,
 \quad |C_q-C_s|\le3n^2F_1|q-s|.
\]
For the first bound, a changed layer has probability at most
$|q-s|\sum_e\Pr(X_e>z)$ and its number of coloops is at most $n$;
integrate and use $\sum_e\E X_e\le nF_1$. For $H$, differences of
the factors $p_e^2$ contribute at most twice the sum of probability
changes, while the nonspanning factors contribute at most $n-1$
times that sum. This proves the other two bounds.
Also $F_q\ge q_-F_1$ and, by concavity,
$|F_q-F_s|\le(F_1/q_-)|q-s|$. Since
$d_*'=\theta/a^2$, both $C_q/F_q-d_*(q)$ and $D_q/F_q-d_*(q)$ are Lipschitz
with constant at most
\[
 \left(\frac3{q_-}+\frac1{q_-^2}+\frac\theta{a_-^2}\right)n^2
 \le14n^2<16n^2.
\]

At an initial stopping point, the failed tests give the desired two
inequalities with $15\varepsilon$ in place of $20\varepsilon$.
In the positive branch the saved left endpoint is certified positive.
If the final right endpoint is interior, it was assigned by a failed
positive test, so its true normalized $C$ ratio is at most
$15\varepsilon$ above $d_*$. Transfer this estimate across the final
interval, losing at most $4\varepsilon$, to obtain the upper inequality
at the returned point. The positive certificate gives the lower one.
The negative branch is symmetric. If the opposite endpoint was never
updated, the returned point is within $\varepsilon/(4N^2)<\delta$
of an endpoint of $J$, and hence lies in the boundary region
$|q-1|\ge3\delta$. This proves the assertion in every case.
\end{proof}

All ratio tests can be performed by cross multiplication. The estimates
share the factor $V_q/m$, and every new parameter is a midpoint of two
old endpoints. Appendix~\ref{sec:combined-implementation} uses these
facts to bound the bit complexity of the search.

\section{The online algorithm and its guarantee}\label{sec:policy}

The selected reference serves two purposes. Its location and its stable
surplus determine whether the virtual continuation can earn the entire
remaining potential. If the algorithm instead takes a threshold step,
the same reference supplies a price for each possible next arrival.

\subsection{Conditions on the parameters}

The three inequalities below correspond to the two ways of entering
the continuation and the single threshold step. Put
\begin{equation}\label{eq:branch-parameters}
 \begin{gathered}
 h_-=\eta-10\varepsilon,\qquad h_+=\eta+10\varepsilon,\qquad
 b=b(q)=a(q)\frac{1+\varepsilon}{1-\varepsilon},\\
 z_*(q)=\frac{h_-^2}{1-d_*(q)+20\varepsilon+h_-},\\
 L_*(q)=1-b+\frac{b}{b+s}
       \left(d_*(q)+20\varepsilon-(1-b)-be^{-s}\right).
 \end{gathered}
\end{equation}
The elementary restrictions~\eqref{eq:basic-parameters} make the
denominator of $z_*$ positive and give $0<b<1$.
In addition to $\omega>0$ and $q(b+s)>1$ throughout $J$, require
\begin{align}
 \Phi_\delta(q,0)&\ge a(q)
       &&\text{if }q\in J,\ |q-1|\ge3\delta,
       \label{eq:condition-guard}\\
 \Phi_\delta(q,z_*(q))&\ge a(q)
       &&\text{for }q\in J,
       \label{eq:condition-stability}\\
 L_*(q)&>bq\bigl(d_*(q)+h_++20\varepsilon\bigr)+2\varepsilon
       &&\text{if }|q-1|\le3\delta.
       \label{eq:condition-threshold}
\end{align}
The first condition covers a reference close to the edge of the
search interval. The second pays for a continuation with large
stable surplus. The third pays for a threshold step when stable
surplus is small. Appendix~\ref{sec:parameters} verifies these
conditions simultaneously.

\subsection{The policy}

If $A$ is the set of accepted elements and $R$ the set of rejected or
discarded elements, the residual matroid is $(M_0/A)\setminus R$.
At each decision, the algorithm computes the reference and all prices
before revealing the next remaining label. The value of $N$ in the error
allowances stays fixed throughout.

\begin{algorithm}[t]
\caption{The matroid prophet-secretary algorithm}
\label{alg:outer}
\textbf{Input:} $M_0=(E_0,\mathcal I_0)$ through a matroid independence
oracle, the explicit input distributions, and the fixed parameters.\\
\textbf{Output:} an independent accepted set $A$.
\begin{enumerate}[leftmargin=*,itemsep=3pt,topsep=5pt]
\item Initialize $N\gets|E_0|$, $A\gets\varnothing$,
$M\gets M_0$, and $q^{\rm in}\gets1$.
\item \textbf{Repeat}:
\begin{enumerate}[label=(\alph*),leftmargin=1.6em,itemsep=3pt,topsep=2pt]
\item Delete every loop from $M$ and ignore its future arrival.
If $E(M)=\varnothing$, or if $X_e=0$ almost surely for every
$e\in E(M)$, reject all remaining arrivals and \textbf{return $A$}.
\item Run the reference search of Section~\ref{sec:combined-search}
on $(M,q^{\rm in})$. Retain its selected $q$, the exact $V_q$, and
the saved estimates $\widehat F,\widehat D,\widehat C,
(\widehat m_e)_{e\in E(M)}$.
\item If $|q-1|\ge3\delta$ or
$(\widehat D-\widehat C)/\widehat F>\eta$, run the virtual
continuation on $M$ with parameter $q$ and fresh randomness.
Let $A'$ be its accepted set. \textbf{Return $A\cup A'$}.
\item For every $e\in E(M)$, set
\[
 p_e\gets a(q)\,
       \frac{\widehat m_e+\varepsilon V_q/N}{1-\varepsilon}.
\]
\item Observe the next remaining element $e$ and its value $X_e$.
If $X_e>p_e$, accept $e$, set $A\gets A\cup\{e\}$ and
$M\gets M/e$. Otherwise reject $e$ and set $M\gets M\setminus e$.
\item Set $q^{\rm in}\gets q$ and repeat.
\end{enumerate}
\end{enumerate}
\end{algorithm}

A loop is ignored when it arrives. Conditional on the history, the
relative order of the remaining nonloops is uniform and their values
retain their original product law. Reference samples are independent
of these future arrivals. A threshold acceptance preserves feasibility
by contraction. At entry to a virtual continuation we use the current
minor and the selected reference, initialize its local accepted set
and record list to be empty, and attach fresh sorted clocks to the
remaining arrivals as in Section~\ref{sec:controller}.

Each threshold step removes one label. There are therefore at most $N$
searches and at most one virtual continuation. A reference passed to
another search always satisfies $|q-1|<3\delta$. When $N=0$, the
algorithm returns the empty set before using $N$ in a denominator.

\subsection{One-step progress}

\begin{lemma}[Local potential bound]\label{lem:local-potential}
Assume~\eqref{eq:basic-parameters}
and~\eqref{eq:condition-guard}--\eqref{eq:condition-threshold},
as well as $\omega>0$ and $q(b+s)>1$ throughout $J$.
At a positive-value residual state, on the event that all queried
batches are accurate, the next action has expected reward plus
continuation potential at least
\[
                        a(q^{\rm in})F_{q^{\rm in}}.
\]
After a threshold step, the potential is evaluated on the child
minor at the inherited selected parameter $q$. A virtual continuation
has zero terminal potential.
\end{lemma}

\begin{proof}
Fix all queried batches at the state and write $q$ for the selected
reference. Abbreviate $a=a(q)$, $F=F_q$, $D=D_q$, and $H=H_q$.
Reference selection gives $aF\ge a(q^{\rm in})F_{q^{\rm in}}$,
so it suffices to pay for $aF$.

If $|q-1|\ge3\delta$, the continuation earns at least $aF$ by
Proposition~\ref{prop:controller}, $Q\ge0$, and~\eqref{eq:condition-guard}.
It remains to consider $|q-1|<3\delta$. Reference selection then gives
\begin{equation}\label{eq:local-stationarity}
             d_*-20\varepsilon\le D/F
                        \le d_*+H/F+20\varepsilon.
\end{equation}
The two ratio estimates used by the search also imply
\[
 \left|\frac{\widehat D-\widehat C}{\widehat F}-\frac HF\right|
                                                    \le10\varepsilon.
\]

\paragraph{Large stable surplus.}
If $(\widehat D-\widehat C)/\widehat F>\eta$, then $H/F>h_->0$.
By Lemma~\ref{lem:stable-overlap}
and~\eqref{eq:local-stationarity},
\[
 \frac QF
 \ge\frac{(H/F)^2}{1-D/F+H/F}
 \ge\frac{(H/F)^2}{1-d_*+20\varepsilon+H/F}
 \ge z_*(q).
\]
The last inequality uses that $x^2/(c+x)$ is increasing for
$x\ge0$ when $c>0$. Since $\Phi_\delta(q,z)$ increases with $z$,
condition~\eqref{eq:condition-stability} and
Proposition~\ref{prop:controller} give reward at least $aF$.

\paragraph{Small stable surplus.}
Otherwise $H/F\le h_+$. Let $b$ be as in
\eqref{eq:branch-parameters}, and define
$S_b=\sum_e\E(Y_e^q-bm_e)^+$.
Write $d=D/F$, $h=H/F$, and $c=b/(b+s)$. We compare the
surplus directly with the benchmark loss. Corollary~\ref{cor:structural-surplus}
gives
\[
 \frac{S_b}{F}-bqd
 \ge 1-b-c\bigl((1-b)+h+be^{-s}\bigr)+(c-bq)d.
\]
Since $q(b+s)>1$, the coefficient $c-bq$ is negative. We may
therefore substitute the upper stationarity bound
$d\le d_*+h+20\varepsilon$. The terms $ch$ cancel, giving
\[
\begin{aligned}
 \frac{S_b}{F}-bqd
 &\ge 1-b+c\bigl(d_*+20\varepsilon-(1-b)-be^{-s}\bigr)
                    -bq(d_*+h+20\varepsilon)\\
 &\ge L_*(q)-bq(d_*+h_++20\varepsilon)>2\varepsilon.
\end{aligned}
\]
The last two inequalities use $h\le h_+$ and
condition~\eqref{eq:condition-threshold}. Thus
\begin{equation}\label{eq:threshold-reference-surplus}
                             S_b>bqD+2\varepsilon F.
\end{equation}

Fix the vector of thresholds $p_e$ determined by the saved batch. Equations~\eqref{eq:combined-upper-price}
and~\eqref{eq:combined-price-sum} give
$p_e\ge am_e$ and
$\sum_e(p_e-bm_e)^+<2\varepsilon F$.
Let $S_X=\sum_e\E(X_e-p_e)^+$. Reference tails satisfy
$\Prb(Y_e^q>z)\le q\Prb(X_e>z)$, and hence
\[
 S_b\le\sum_e\E(Y_e^q-p_e)^++\sum_e(p_e-bm_e)^+
                         <qS_X+2\varepsilon F.
\]
Together with~\eqref{eq:threshold-reference-surplus}, this yields
$S_X>bD\ge aD$.

For a fixed next label $e$, the difference between the deletion
and contraction potentials is
\[
                    aF_q(M\setminus e)-aF_q(M/e)=am_e.
\]
Since $p_e\ge am_e$, accepting when $X_e>p_e$ improves on deletion
by at least $(X_e-p_e)^+$. Averaging over the next uniform label
therefore gives expected reward plus child potential at least
\[
                aF+\frac1n(S_X-aD)\ge aF.
\]
This child potential is exactly the incoming potential for the
next reference search. All branches satisfy the claimed bound.
\end{proof}

\subsection{Charging statistical errors}

\begin{proposition}[Guarantee with exact random primitives]
\label{prop:ideal-guarantee}
Under the stated parameter conditions, the ideal policy satisfies
\[
                         \E[\ALG]\ge(1-\theta-\rho)F_0.
\]
Its expected number of matroid independence queries is $O(N^4)$.
\end{proposition}

\begin{proof}
If $F_0=0$, the policy stops and the claim is immediate.
Otherwise use the potential $\mathcal P(M,q)=a(q)F_q(M)$.
Initially it is $(1-\theta)F_0$. For every residual minor,
$F_1(M)\le F_0$, so Lemma~\ref{lem:reference-derivative}
and~\eqref{eq:basic-parameters} imply
\[
       0\le\mathcal P(M,q)
       \le\bigl(\max_{u\in J}a(u)\bigr)q_+F_1(M)\le F_0.
\]
Conditionally at each computation state, all queried batches are
accurate with probability at least $1-\rho/N$, by
Lemma~\ref{lem:combined-confidence}. On that event,
Lemma~\ref{lem:local-potential} gives nonnegative expected drift
of reward plus potential. On the complementary event, rewards and
subsequent potentials remain nonnegative, so the expected decrease
is at most $F_0$.

Treat a virtual continuation as one final epoch and append empty
epochs to obtain exactly $N$ epochs. Summing conditional drift
bounds loses at most $N(\rho/N)F_0=\rho F_0$.
No union bound over possible residual minors is needed.

There are at most $NKm=O(N^2\log^2(N+2))$ layer samples.
Their total query cost is $O(N^3\log^3(N+2))$.
Loop detection and actual decisions use $O(N^2)$ further queries.
The one possible virtual continuation costs $O(N^4)$ in expectation.
Since $\log^3(N+2)=O(N)$, the total is $O(N^4)$.
\end{proof}

Appendix~\ref{sec:combined-implementation} converts this policy to
a bounded-bit algorithm at an additional loss of at most $\rho F_0$.
The resulting competitive ratio is $1-\theta-2\rho$.

\section{The sharp bound for XOS valuations}
\label{sec:xos-upper}

We now turn from selecting elements of a matroid to allocating disjoint
bundles to buyers. For XOS valuations, Ehsani et al.~\cite{EHKS2018}
give a $1-1/e$ guarantee. We prove the matching upper bound even when
every buyer draws her valuation from the same distribution. Our instance
has a simple description. Each buyer values the items she receives in
one part of a partition, and her type specifies which parts she may
use. The prophet can usually choose a single partition that serves
all buyers. An online algorithm must allocate each bundle before it
knows which partition will serve the later buyers.

\subsection{A family of partitions}
\label{subsec:xos-partitions}

Fix $n\ge2$ buyers and a finite index set $\mathcal T$. An item is a
function from $\mathcal T$ to $[n]$, so the item set is
$M=[n]^{\mathcal T}$ and $m=|M|=n^{|\mathcal T|}$. Each coordinate
$s\in\mathcal T$ partitions $M$ into the $n$ equal parts
\[
 P_s^r=\{x\in M:x(s)=r\},\qquad r\in[n].
\]
All these partitions are public. Within a partition the $n$ parts are
disjoint. Across partitions their overlaps are given by the product rule.

\begin{lemma}[Counting partition parts]
\label{lem:xos-partitions}
For distinct $s_1,\ldots,s_k\in\mathcal T$ and arbitrary
$r_1,\ldots,r_k\in[n]$,
\[
 \left|\bigcup_{i=1}^k P_{s_i}^{r_i}\right|
 =m\left(1-\left(1-\frac1n\right)^k\right).
\]
\end{lemma}
\begin{proof}
Outside the union each of the $k$ specified coordinates has $n-1$
possible values. Each of the other coordinates has $n$ possibilities.
\end{proof}

\subsection{An i.i.d. distribution of buyer types}
\label{subsec:xos-iid}

Fix $L\ge n$ and index the partitions by
$\mathcal T=\binom{[L]}{n}$. Give each buyer an independent uniform
label in $[L]$. For any partition index $s$ containing her label $a$,
she may use part $r_s(a)$, where $r_s(a)$ is the position of $a$ in
the increasing order on $s$. Her valuation for a bundle $S$ is
\begin{equation}
\label{eq:xos-valuation}
 v^a(S)=\max_{s\in\mathcal T:\,a\in s}|S\cap P_s^{r_s(a)}|.
\end{equation}
This is the maximum of additive clauses with coefficients in
$\{0,1\}$. In particular,
\begin{equation}
\label{eq:xos-cardinality}
 v^a(S)\le |S|,\qquad v^a(M)=m/n.
\end{equation}

\paragraph{The prophet.}
If the $n$ labels are distinct, their set is a partition index $s$.
Allocating the part $P_s^{r_s(a)}$ to each buyer with label $a$
assigns every item exactly once, for welfare $m$. A pair of labels
coincides with probability $1/L$. Thus
\begin{equation}
\label{eq:xos-prophet}
 \E[\OPT]\ge m\left(1-\binom{n}{2}/L\right).
\end{equation}

\paragraph{The online allocation.}
Reveal the entire permutation of buyer identities in advance and the
label of each buyer at her arrival. Conditional on the permutation
and the algorithm's independent random tape, the labels
$A_1,\ldots,A_n$ in arrival order remain independent and uniform.
When the algorithm gives $S_t$ to the buyer at position $t$, fix the
first maximizing index $W_t$ in a public ordering of the clauses
in~\eqref{eq:xos-valuation}. Then
\begin{equation}
\label{eq:xos-witness}
 A_t\in W_t,\qquad
 v^{A_t}(S_t)=|S_t\cap P_{W_t}^{r_{W_t}(A_t)}|.
\end{equation}
We can fix this witness at the moment of allocation, even if several
clauses attain the same value. For $t<u$, the witnesses can coincide
only if the future label $A_u$ lies in the already fixed $n$-set $W_t$.
Conditional on the history through $t$, that event has probability
$n/L$. A union bound gives
\begin{equation}
\label{eq:xos-witness-collision}
 \Pr[\exists\,t<u:\ W_t=W_u]\le \binom{n}{2}\frac{n}{L}.
\end{equation}

When all witnesses are distinct, the $n$ credited sets in
\eqref{eq:xos-witness} are disjoint subsets of their respective
partition parts. Hence, by Lemma~\ref{lem:xos-partitions}, their total
size is at most $m c_n$, where $c_n=1-(1-1/n)^n$. This counting
identity holds for every choice of distinct indices, including the
ones selected by the algorithm's past decisions. On a collision the
welfare is at most $m$ by~\eqref{eq:xos-cardinality}. Averaging over
the permutation and the random tape gives
\begin{equation}
\label{eq:xos-online}
 \E[\ALG]\le m\left(c_n+n\binom{n}{2}/L\right).
\end{equation}

\begin{proof}[Proof of Theorem~\ref{thm:xos-upper}]
For $L>\binom n2$, dividing~\eqref{eq:xos-online}
by the lower bound in~\eqref{eq:xos-prophet} gives
\begin{equation}
\label{eq:xos-ratio}
 \frac{\E[\ALG]}{\E[\OPT]}
 \le \frac{c_n+n\binom{n}{2}/L}{1-\binom{n}{2}/L}.
\end{equation}
Given $\xi>0$, choose $n$ so that $c_n<1-1/e+\xi/2$ and then
$L$ so that the right-hand side is below $1-1/e+\xi$.
These choices work for every online algorithm.
\end{proof}

The bound allows unlimited computation and full knowledge of the input
law. Indeed the list of clauses, over all $L$ possible types,
has $L\binom{L-1}{n-1}=n|\mathcal T|$ clause occurrences. Listing the $m$ binary coefficients of each takes
$nm\log_n m$ bits. The uncertainty is about which types will arrive.

\begin{corollary}\label{cor:xos-items}
For $0<\xi<1/4$, the instance in Theorem~\ref{thm:xos-upper} can have
$O(\xi^{-4}\log(1/\xi))$ items.
\end{corollary}
\begin{proof}
Set $c=1-1/e$, $n=1+\lceil8/\xi\rceil$, and $L=\lceil8n^3/\xi\rceil$.
Sample $m'$ items independently from $M$, treating repeated samples as
distinct items. Restrict each clause to these items. For any fixed
choice of $n$ distinct partition indices and one part from each,
the fraction of sampled items in their union is the mean of $m'$
independent Bernoulli variables of mean $c_n$. Let $\zeta=\xi/8$ and
take $m'=\lceil (1+n\log(n|\mathcal T|))/(2\zeta^2)\rceil$.
Hoeffding's inequality and a union bound over at most
$(n|\mathcal T|)^n$ choices give a realization in which every such
union covers at most a $c_n+\zeta$ fraction. Fix this realization
and make it public.

Every partition still covers all $m'$ items, so the prophet earns $m'$
whenever the labels are distinct. The collision bound~\eqref{eq:xos-witness-collision} still holds because each witness is indexed by a set of $n$ types.
Consequently~\eqref{eq:xos-ratio} holds with $c_n+\zeta$ in place of
$c_n$. The power series for $\log(1-1/n)$ gives
$n\log(1-1/n)\ge-1-1/(2(n-1))$, whence
$c_n-c\le 1/(2e(n-1))<\xi/16$. Also
$n\binom{n}{2}/L<\xi/16$. The ratio is therefore at most
$(c+\xi/4)/(1-\xi/16)<c+\xi$. Finally,
$|\mathcal T|\le L^n$ gives the stated bound on $m'$.
The number of clauses per valuation can still be exponential in $n$.
\end{proof}

\subsection{Relation to coverage welfare}
\label{subsec:xos-coverage}

The upper bound also follows from Udwani and Wajc's theorem for online
coverage welfare under known i.i.d. item arrivals~\cite[Theorem~1.2]{UW2026}.
We give the reduction to clarify the change of arrival model. It uses
the coverage-to-XOS construction of Chawla and Dang~\cite[Appendix~D]{CD2026},
who apply it to allocation with free disposal.

In the coverage problem there are $k$ fixed buyers. Buyer $a$ has
a universe $U_a$ with nonnegative weights $w_{a,u}$, and an arriving
item of type $e$ covers $C_{a,e}\subseteq U_a$ if assigned to her.
Her value is the total weight of the union of sets covered by her
assigned items. Take disjoint copies of the universes. In the auction
their elements become the items. Each arrival of type $e$ becomes a
buyer with valuation
\begin{equation}
\label{eq:xos-transposition}
 v_e(S)=\max_{a\in[k]}\sum_{u\in S\cap C_{a,e}}w_{a,u}.
\end{equation}
To simulate an auction algorithm in the coverage problem, assign the
arriving item to a maximizing clause $a$ for the bundle $S$ it
allocates. All elements counted in that clause are then covered by
buyer $a$. Because the auction bundles are disjoint, these elements
are charged at most once across arrivals. The coverage welfare
therefore dominates the auction welfare. Conversely, whenever a
coverage algorithm assigns $e$ to $a$, give the corresponding
auction buyer the elements of $C_{a,e}$ still unallocated. They are
exactly the elements newly covered by this assignment, and the
bundle earns their total weight. Both simulations run online.
Applied offline they give equality of the optima. Finally, i.i.d.\ item types induce i.i.d.\ buyer valuations, whose joint law is
preserved by an independent random permutation. Thus the
information-theoretic coverage upper bound transfers to this auction
model.

\section{AI Use Statement}
This work resulted from extensive interaction between the authors and OpenAI
Astra, with the authors providing suggestions and critically checking Astra's responses through
iterative exchanges. OpenAI Astra was also used in writing and revising the manuscript through
detailed prompts from the authors, with Gemini Argon assisting in some parts. The authors directed the process and evaluated the proposed ideas and arguments. A supplementary Lean archive is available on GitHub [\url{https://github.com/hajiaghayi/Matroid-Prophet-Secretary-XOS}].

\bibliographystyle{alpha}
\bibliography{prophet_secretary}

\clearpage
\appendix

\section{Implementing the common-law rule}\label{iid:sec:implementation}

Let $n=|E(M)|\ge1$. An independence query to $M$ costs $T_I$, and a
query returning $F(x-)$ and $F(x)$ costs $T_F$. We first count arithmetic
operations, comparisons, and uniform draws on $[0,1]$ at unit cost.
We give the finite-bit version below. Let $I$ be the accepted set. To
test whether $A\subseteq E(N)\setminus\{e\}$ spans $e$, greedily
extend $I$ through $A$, then test whether the resulting independent set
can be augmented by $e$. The test takes at most $|A|+1$ queries.
Applying it to the empty set and the whole remaining set detects loops
and coloops.

\begin{algorithm}[H]
\caption{The sampled rule for identically distributed weights}\label{iid:alg:sample}
\begin{enumerate}[label=\arabic*.,leftmargin=*]
\item Reject a loop and accept a coloop. For every other arrival $(e,x)$,
draw its upper quantile $u=1-F(x)+V(F(x)-F(x-))$, using a fresh uniform $V$.
\item Draw $m$ independent subsets of $E(N)\setminus\{e\}$ by including
each label with probability $u$. Let $H$ be the number of subsets that
fail to span $e$.
\item Accept if $H/m\ge e^{-\eta_0}$. Contract an accepted element and
delete a rejected element.
\end{enumerate}
\end{algorithm}

\begin{proposition}\label{iid:prop:implementation}
For an absolute constant $C$, using
$m=\lceil Cn^2\lceil\log_2(2n)\rceil\rceil$ trials per arrival in
Algorithm~\ref{iid:alg:sample} gives competitive ratio $c_0$. The
running time is $O(n^4\log(2n)(1+T_I)+nT_F)$.
\end{proposition}

The sampling error would cost a small additive term if we used only
the guarantee in Theorem~\ref{iid:thm:band}. On a ground set of size $n$,
however, the comparison has slack of order $1/n$. We prove this first
and then use the slack to pay for sampling.

\begin{proof}
Let $g(a)=e^{-1-a}/a$. Consider a deterministic cutoff $q$ calibrated
at $S(q)=e^{-a}$ with $\eta_0/2\le a\le3\eta_0/2$, for a label that is
neither a loop nor a coloop. There are absolute constants $K,B>0$ with
\begin{equation}
 \frac{D(v)}{\min(q,v)}
 \le 1+\eta_0+K|a-\eta_0|-B/n \qquad(0<v\le1).           \label{iid:eq:slack}
\end{equation}
Indeed, the event of sampling no other label gives $S(q)\ge(1-q)^{n-1}$.
Thus $q\ge1-e^{-a/(n-1)}\ge b/n$ for $b=1-e^{-\eta_0/2}$. Set
$d=1-q/\tau(q)$. Since $\tau(q)\ge q+q^2/2$, we have $d\ge q/3$.
For $v\le q$, Lemma~\ref{iid:lem:shape} gives
$1-S(v)\le a\tau(v)/\tau(q)$. By
$(1-v/q)\tau(v)/v\le(1-v)\tau(v)/v\le1$, it follows that
\[
 D(v)/v-1\le a(1-d)\le a-\eta_0 q/6.
\]
For $v=qx\ge q$, the power series of $\tau$ yields
$\tau(qx)/\tau(q)\ge (1-d)x+dx^2$, first for $v<1$ and then by a
limit. Applying Lemma~\ref{iid:lem:shape} in the other direction gives
\[
 D(v)/q-1\le (x-1)e^{-ax-adx(x-1)}.
\]
For $1\le x\le1+\eta_0/2$ this is at most $\eta_0/2$. For $x>1+\eta_0/2$
it is at most $g(a)e^{-\eta_0^2q/12}$. Concavity gives
$1-e^{-\eta_0^2q/12}\ge q(1-e^{-\eta_0^2/12})$. Since $g(\eta_0)=\eta_0$,
these inequalities prove~\eqref{iid:eq:slack}, for instance with
\begin{align*}
 K&=1+\max_{a\in[\eta_0/2,3\eta_0/2]}|g'(a)|,\\
 B&=\min\{\eta_0/2,\,\eta_0 b/6,\,b g(3\eta_0/2)(1-e^{-\eta_0^2/12})\}.
\end{align*}

To analyze the sample, give each label an independent uniform priority
in each trial and include it whenever its priority is at most $u$.
As $u$ grows, the number of nonspanning trials can only decrease.
Thus, conditioned on these priorities, the sampled decision is a cutoff
$\widehat q$ independent of the new weight and its tie breaker. Whether
an interval endpoint belongs to the acceptance region is immaterial
under continuous quantiles.

Put $h=B/(2Kn)$ and define $q_-<q_+$ by
$S(q_-)=e^{-\eta_0+h}$ and $S(q_+)=e^{-\eta_0-h}$.
At both endpoints the population mean differs from
$\beta=e^{-\eta_0}$ by at least $\rho=\beta(1-e^{-h})=\Theta(1/n)$.
Hoeffding's inequality~\cite{hoeffding} and monotonicity give
\[
    \Pr(\widehat q\notin[q_-,q_+])\le2e^{-2m\rho^2}.
\]
This holds conditionally on the current minor. Choose
$\lambda=c_0^2 B/4$ and take $C$ large enough that the right-hand
side is at most $\lambda/n^2$. For every cutoff in the interval,
\eqref{iid:eq:slack} pays a fraction of each rank drop at least
\[
 (1+\eta_0-B/(2n))^{-1}\ge c_0+2\lambda/n.
\]
This fraction also works for loops and coloops since
$B\le\eta_0/2$ and $1-c_0=c_0\eta_0$.

For the analysis only, replace a cutoff outside the interval by the
exact one. The resulting policy earns at least
$(c_0+2\lambda/n)P_M(F)$ by the band argument. Couple it to
Algorithm~\ref{iid:alg:sample} up to the first replacement. Even
conditionally on \emph{all} weights, tie breakers, and the arrival order,
the unused priorities at each step are fresh. The probability of a
replacement is therefore at most $n(\lambda/n^2)=\lambda/n$.
On this event the difference in payoffs is bounded by the realized
offline optimum. Averaging gives at least $(c_0+\lambda/n)P_M(F)$
for the sampled algorithm. Each trial uses $O(n)$ elementary operations
and $O(n)$ independence queries using a mutable set representation, which
gives the claimed running time.
\end{proof}

\begin{corollary}[Explicit rational distributions]\label{iid:cor:bits}
If the common law is specified by rational weights and probabilities on a
finite support, the ratio $c_0$ is achievable with polynomial worst-case
bit complexity and $O(n^4\log(2n))$ independence queries.
\end{corollary}
\begin{proof}
Choose a dyadic $\widetilde\beta$ within $\rho/2$ of $e^{-\eta_0}$.
The preceding proof still gives $c_0+\lambda/n$ for the real-arithmetic
algorithm using $\widetilde\beta$. Replace $\rho$ by $\rho/2$ in
the Hoeffding bound and increase $C$. Computing this dyadic requires
$O(\log(2n))$ bits of accuracy. For instance, bisect the strictly
increasing function $a^2-e^{-1-a}$ on $[1/4,1/2]$, using the Taylor
series for the exponential with a remainder bound. Its derivative is
at least $1/2$ on this interval, so a function evaluation within
$\epsilon$ of zero locates the root within $2\epsilon$. This gives a
certified approximation in time polynomial in the requested precision.

Replace every uniform priority and tie breaker by its first $L$ binary
digits. Couple both versions by truncating the continuous draws.
The quantile $u$ changes by at most $2^{-L}$ even conditionally on the
entire vector of weights and continuous tie breakers. A comparison between
a priority and $u$ can then change only if the original priority is
within $2^{1-L}$ of $u$. Each fresh priority is uniform, independently
of this conditioning, so the probability of such an event is at most
$2^{2-L}$. Up to the first disagreement there are at most $n^2m$
comparisons. Choose $L=O(\log(2n))$ with
$2^{2-L}n^2m\le\lambda/n$. The conditional probability of a disagreement
is at most $\lambda/n$, and its cost is at most the realized offline
optimum. This proves the ratio. The cumulative masses of the explicit
law and all truncated comparisons can be computed exactly using
rationals of polynomial bit length.
\end{proof}

\section{Optimal quantile cutoffs in uniform random order}\label{iid:sec:flows}

The common-law algorithm in Section~\ref{iid:sec:band} works for every
fixed arrival order independent of the draws. With a uniform permutation we can characterize the best
guarantee for each individual matroid without imposing a local comparison.
Fix a matroid $M$ of positive rank on $n$ labels, and let $V_M(F)$ be the value of its
optimal online policy in uniform random order. Write
\[
       \gamma(M)=\inf_{F:P_M(F)>0}\frac{V_M(F)}{P_M(F)},
\]
where $F$ ranges over nonnegative common laws of finite mean. The policy
inside the infimum may use the entire law. The next theorem shows that
one table of quantile cutoffs, chosen knowing only the matroid,
achieves $\gamma(M)$ simultaneously for all common laws.

\paragraph{A flow on minors.}
Make a vertex for each residual labeled minor $N$. The root $M$ has mass
one. If a nonempty $N$ receives mass $y_N$, the mass at which $e$ arrives
next is $b_{Ne}=y_N/|E(N)|$. Send $a_{Ne}$ units of this mass to $N/e$
and $b_{Ne}-a_{Ne}$ to $N\setminus e$, with $0\le a_{Ne}\le b_{Ne}$
and $a_{Ne}=0$ for loops. Sum the incoming masses at each vertex,
including both arcs if they reach the same child. The graph is acyclic,
and the feasible flows form a compact polytope $\mathcal D_M$. A flow
$D$ prescribes a cutoff $q_{Ne}=a_{Ne}/b_{Ne}$ at every positive-mass
arrival. Put
\[
                  T_D(v)=\sum_{N,e}\min(a_{Ne},v b_{Ne}).
\]

\begin{theorem}[Simultaneous quantile cutoffs]\label{iid:thm:flow}
In uniform random order,
\[
 \gamma(M)=\max_{D\in\mathcal D_M}\inf_{0<v\le1}
                    \frac{T_D(v)}{R_M(v)}.
\]
The quantile cutoffs of a maximizing flow attain this ratio for every
common law.
\end{theorem}

\begin{proof}
Any adaptive policy induces a feasible flow. Conditional on the past,
the next label is uniform on the unseen labels, and the arriving
quantile is fresh. Of the total mass $b_{Ne}$, at most $a_{Ne}$ is
accepted and at most $v b_{Ne}$ belongs to the top $v$ quantile band.
Accepting its best $q_{Ne}$ fraction attains the minimum of the two.
This change preserves the masses of both children, so by induction it
realizes the entire flow and earns exactly $T_D(v)$ on every band.
Zero-mass states can use arbitrary feasible decisions. It follows by
integration over weight levels that any flow satisfying
$T_D(v)\ge c R_M(v)$ for all $v$ is $c$-competitive on every law.

It remains to find such a flow for $c=\gamma(M)$. Fix finitely many
positive endpoints $v_1,\ldots,v_k$. The set of vectors dominated
coordinatewise by $(T_D(v_i))_i$ for some $D$ is closed by compactness
and convex by concavity of each $T_D(v_i)$ in $D$. If this set excluded
$(cR_M(v_i))_i$, strict separation from that point by a hyperplane
would give coefficients $\lambda_i\ge0$, not all zero, with
\[
  \max_{D\in\mathcal D_M}\sum_i\lambda_iT_D(v_i)
    < c\sum_i\lambda_i R_M(v_i).
\]
Nonnegativity follows since the set is closed under decreasing any
coordinate. Give a label of upper quantile $u$ the weight
$\sum_i\lambda_i\mathbf 1_{\{u\le v_i\}}$. Its expected prophet value
is the sum on the right without $c$, by the greedy algorithm. The sum
on the left bounds the value of every policy, contradicting the
definition of $\gamma(M)$. Thus every finite collection of the closed
constraints $T_D(v)\ge cR_M(v)$ admits a common flow. Compactness gives
a flow satisfying all of them.
\end{proof}

For example, the deficit $R_M(v)-T_D(v)$ is the sum over nonloop
arrivals of positive mass of
\[
 b_{Ne}\bigl[(q_{Ne}-v)_+(1-S_e^N(v))+(v-q_{Ne})_+S_e^N(v)\bigr].
\]
This follows by telescoping~\eqref{iid:eq:drop} along the flow. The
characterization therefore permits a loss at one arrival to be offset
by a smaller loss at another.

\paragraph{Finite approximation.}
For $0<\delta\le1$, start at $v_1=\delta/n$, increase the endpoints
by a factor $1+\delta$, and append $1$ as the last endpoint. On this
grid maximize $c$ subject to the flow constraints and
\[
 \sum_{N,e}t_{Ne,i}\ge cR_M(v_i),\qquad
 0\le t_{Ne,i}\le a_{Ne},\qquad t_{Ne,i}\le v_i b_{Ne}.
\]
The resulting linear program has at most $3^n$ minor vertices and
$O(1+\delta^{-1}\log(n/\delta))$ endpoints. Its cutoffs attain
$\gamma(M)/(1+\delta)$ on every law.

Indeed, let $c$ be the grid optimum, so $c\ge\gamma(M)$. Both $T_D$ and
$R_M$ are nondecreasing concave functions of $v$ that vanish at zero.
For the latter, differentiating the Bernoulli expectation gives
$R_M'(v)=\sum_e S_e^M(v)$, which is nonincreasing. Between successive
endpoints use monotonicity of $T_D$ and
$R_M(v_{i+1})\le(1+\delta)R_M(v_i)$. Below $v_1$, let $n_0$ be
the number of nonloops. The first two inclusion--exclusion terms for
sampling at least one nonloop give
$R_M(v_1)\ge n_0v_1(1-\delta/2)$. Now
$T_D(v)\ge vT_D(v_1)/v_1$, $R_M(v)\le n_0v$, and
$1-\delta/2\ge 1/(1+\delta)$.

\subsection{Decision prices on random restrictions}\label{iid:sec:restriction-prices}

The flow follows minors under a particular policy. A different identity
compares the optimal policy and the prophet on the same random
restrictions. Let $\mu=\E X$ and $\phi(t)=\E(X-t)_+$. For a nonloop
$e$ of a minor $N$, define
\[
                 w_e=V_{N\setminus e}(F)-V_{N/e}(F)
\]
and let $Z_e$ be the difference between the \emph{realized} offline
optima in the same two minors, with identical draws on their common
labels. The optimal online rule compares $X_e$ with $w_e$. The prophet
can compare $X_e$ with $Z_e$ after seeing the other weights. Put $w_e=Z_e=0$
on loops.

For $0\le s\le\ell\le n$, take disjoint uniform sets of $s$
contractions and $\ell-s$ deletions. Write $N_{\ell s}$ for the
resulting minor and $v_{\ell s}=\E V_{N_{\ell s}}(F)$. When $\ell<n$,
let $e$ be a uniform remaining label and $A_e$ the event that $e$ is
neither a loop nor a coloop in this minor.

\begin{theorem}[Prices in random minors]\label{iid:thm:slices}
For the uniform label in $N_{\ell s}$,
\begin{align*}
 \E w_e&=v_{\ell+1,s}-v_{\ell+1,s+1},\\
 v_{\ell s}-v_{\ell+1,s}
   &=\Pr(e\text{ is a coloop})\mu+\E[\mathbf 1_{A_e}\phi(w_e)].
\end{align*}
In particular, let $N_k$ be the restriction to a uniform $k$-element
set. With the expectations in the sums taken over $N_k$ and its uniform
label,
\begin{align*}
 V_M(F)&=r(M)\mu+\sum_{k=1}^n\E[\mathbf 1_{A_e}\phi(w_e)],\\
 P_M(F)&=r(M)\mu+\sum_{k=1}^n\E[\mathbf 1_{A_e}\phi(Z_e)].
\end{align*}
\end{theorem}
\begin{proof}
Deleting the uniform label gives the law of $N_{\ell+1,s}$. Contracting
it gives the law of $N_{\ell+1,s+1}$. The two minors coincide on a loop,
so averaging the definition of $w_e$ proves the first identity.
By the Bellman equation, on a nonloop the optimal rule gains
$\phi(w_e)$ in expectation over the deletion continuation. A loop
adds zero. A coloop adds $\mu$ since its two continuation minors agree.
Averaging proves the second identity. For the prophet, condition on
the other weights. The gain from a nonloop is now $\phi(Z_e)$, again
equal to $\mu$ on a coloop.

Set $s=0$ and add these identities along successive uniform deletions.
The continuation values telescope. The rank drops by one precisely
when the deleted label is a coloop, so the sum of the coloop
probabilities is $r(M)$. This gives both decompositions.
\end{proof}

\section{Finite precision and polynomial running time}
\label{sec:combined-implementation}

We now implement the ideal policy with a deterministic bound on all
random-bit and arithmetic work. The distributions are explicit lists of
nonnegative rational values and rational probabilities. Let $L$ be
their total binary encoding length and $s_{\rm in}$ the number of
support entries. Zero-probability entries may be removed. The
parameters are fixed rational constants; choose $\delta$ dyadic, as
in Appendix~\ref{sec:parameters}.

The statistical estimates cost at most $\rho F_0$ in expected reward.
We give the implementation an additional budget of $\rho F_0$.
A uniform coupling conditional on the actual values will let us
charge an implementation discrepancy to the realized offline optimum.

\subsection{Rational encodings}

Let $D_X$ and $D_v$ be the products of the input probability and value
denominators, respectively. Their bit lengths are $O(L)$. At a queried
reference $q=u/w$, every reference tail and atom probability has
denominator dividing $d_p=wD_X$. Every joint layer weight can be placed
over the common denominator $D_vd_p^n$, so
\[
 V_q=\frac{v}{D_vd_p^n}
\]
for an integer $v$ of polynomial bit length. Thus the joint sampling
table has at most $Ns_{\rm in}$ entries, all of polynomial encoding
length. Very small probabilities do not create a dependence on their
reciprocals.

The averages can be accumulated using integer counters:
\[
 \widehat F=\frac{V_q}{m}k_F,\quad
 \widehat D=\frac{V_q}{m}k_D,\quad
 \widehat m_e=\frac{V_q}{m}k_e,\quad
 \widehat C=\frac{V_q}{md_p}k_C.
\]
Here $k_F,k_D\le mN$, $k_e\le m$, and $k_C\le mNd_p$.
For the last counter, a sampled coloop contributes the integer
$d_p(1-p_{e,j})$. Thus averaging $m$ samples adds only $O(\log m)$ bits.
The affine discount and thresholds have polynomial encoding length.

If the denominator of $\delta$ is $2^d$, every parameter formed by
at most $NK$ midpoint operations has denominator dividing
$2^{d+NK}$. Thus its encoding length is $O(d+NK)$, independently of
the numerical estimates. The input distributions on surviving labels
never change. Thus the encodings have polynomial length over the entire run.

\subsection{A bounded categorical primitive}

Let $S=NKm$, which bounds all layer samples on every execution. Set
\[
 T=\left\lceil\log_2\frac{8N}{\rho}\right\rceil,
 \qquad
 C_{\rm cat}=S(N+2)+N^2(1+T)+2N+NT,
 \qquad J_0=Ns_{\rm in}+N+T+3.
\]
Choose $G=2^{\ell_{\rm bit}}$ so that
\begin{equation}\label{eq:combined-bit-budget}
 G\delta\in\mathbb Z,
 \qquad G\ge\frac{2(4N^2+C_{\rm cat}J_0)}\rho.
\end{equation}
Every auxiliary draw uses exactly $\ell_{\rm bit}$ fair bits to obtain an integer
$J$ uniform on $\{0,\ldots,G-1\}$.

For rational category probabilities $p_1,\ldots,p_k$, form their
cumulative sums and return the first $i$ for which
$\sum_{j\le i}p_j>J/G$. This uses exact rational comparisons.
Zero-probability categories are never returned. The primitive covers
reference coordinates, layer choices, Bernoulli screens, and probe
prefixes.

For an arrival in the early phase of the continuation, draw the probe count from
$\{0,\ldots,T\}$ with probabilities proportional to the rational
weights $2^k/k!$. These are the probabilities of a Poisson random
variable of mean two conditioned to be at most $T$. They may be
represented by the integer weights $2^kT!/k!$. The implementation
therefore requires neither exact exponential evaluations nor an
unbounded count sampler.

Arrival clocks use the one-based grid $(J+1)/G$: draw one clock for
each residual label, sort them, and assign them in order to the actual
arrivals. For a raw probe use a fresh zero-based draw and set
$u=2qJ/G$; test the corresponding prefix only when $u<s$, where $s$
is the current arrival clock. Keep repeated arrival clocks. Candidate
realizations replay all preceding records in their append order. A probe
replays the records whose clocks are at most $u$, also in append order.
Every candidate basis therefore contains all preceding acceptances,
even on executions with equal clocks, and feasibility is unconditional.

\begin{lemma}[Categorical approximation]\label{lem:combined-categorical}
An exact categorical draw with $k$ categories and the bounded-bit draw
can be coupled to disagree with probability at most $(k-1)/G$.
For ordered nonnegative integer categories the bounded-bit draw is no
larger than the exact draw under this coupling.
\end{lemma}
\begin{proof}
Use the same uniform $U\in[0,1)$ and let $J=\lfloor GU\rfloor$.
A discrepancy requires an interior cumulative boundary between $J/G$
and $U$. Each of the $k-1$ boundaries contributes probability at most
$1/G$. The ordering assertion follows because $J/G\le U$.
\end{proof}

\begin{lemma}[Uniform implementation error]\label{lem:combined-bit-coupling}
For every fixed actual value vector in the positive-probability input
supports and every fixed actual arrival permutation, the ideal and
finite-bit algorithms can be coupled to make identical acceptance
decisions with probability at least $1-\rho$. Their expected reward
difference is consequently at most $\rho F_0$.
\end{lemma}

\begin{proof}
First round continuous arrival clocks upward to the grid, retaining
exact categorical draws and Poisson counts. The probability of a
clock collision is at most $n(n-1)/(2G)$; declare a coupling failure
on this event. Otherwise the clock order is preserved. Since the
switching time $1-\delta$ is a grid point, rounding also preserves
the early or late phase.

For a raw early probe, list the preceding accepted times as
$\tau_1\le\cdots\le\tau_k\le s$. Its possible prefix categories
have cumulative boundaries
\[
 \frac{\tau_1}{2q},\ldots,\frac{\tau_k}{2q},\frac{s}{2q},1.
\]
The strict inverse-CDF convention gives exactly the rule $u<s$ and
replay of records at times at most $u$. Rounding a time upward changes
each boundary by at most $1/(2qG)\le1/G$. There are at most $n$
boundaries, so a probe changes category with conditional probability
at most $n/G$. A Poisson count has mean two. Summing over at most $n$
arrivals and including the collision event bounds this first coupling
error by $4N^2/G$. Before a discrepancy, candidate realizations and basis
replays can be coupled identically. The bound holds at every possible
history at which the continuation is entered and for every fixed actual vector and order.

Next replace each Poisson count $Z$ by its law conditioned on $Z\le T$.
The moment bound
\[
 \Pr(Z>T)\le\frac{\E2^Z}{2^{T+1}}
 =\frac{e^2}{2^{T+1}}<4\,2^{-T}
\]
gives error at most $4N2^{-T}\le\rho/2$ over the run. Couple counts
identically when $Z\le T$; otherwise generate a fresh conditional
count and declare failure. This has the required conditional marginal.

Finally replace all rational categorical draws by the bounded-bit
primitive. There are at most $C_{\rm cat}$ draws: each layer uses at
most $N+2$; the at most $N(1+T)$ reference realizations used by the continuation use at most $N$
coordinate draws each; and the remaining draws comprise at most $N$
screens, $N$ counts, and $NT$ prefixes. Every category list has size at
most $J_0$. Until the first discrepancy, histories and the next
categorical laws coincide. A conditional union bound and
Lemma~\ref{lem:combined-categorical} give error at most
$C_{\rm cat}J_0/G$. Combining the three steps with
\eqref{eq:combined-bit-budget} proves total error at most $\rho$.

On the failed coupling, the ideal algorithm's possible advantage is
at most the realized integral optimum: both algorithms are feasible
and have nonnegative reward. The failure bound is uniform conditional
on the actual values. Averaging therefore loses at most $\rho F_0$,
with no range restriction on those values.
\end{proof}

\subsection{Running time}

Every arithmetic operand has polynomial bit length, every categorical
draw uses a fixed number of bits, and each probe count is at most $T$.
The Poisson integer weights have $O(T\log(T+2))$ bits. Thus all loops
and all arithmetic work have polynomial worst-case bounds in $N,L$.
For fixed parameters, $K,T=O(\log(N+2))$, $m=O(N\log(N+2))$,
and $\ell_{\rm bit}=O(\log(N+2)+\log(s_{\rm in}+1))$.

The outer policy uses $O(N^3\log^3(N+2))$ matroid independence queries.
A fresh reference realization requires one greedy basis computation and
at most $n$ exchange replays. The canonical shrinking scan costs
$O(n^2)$ queries per replay, so a reference realization costs $O(n^3)$ queries.
There is at most one continuation. The bounded-bit count
is no larger, in the categorical coupling, than a Poisson count
conditioned on being at most $T$; its mean is at most two. Thus the
continuation algorithm evaluates $O(N)$ reference realizations in expectation and at most
$N(1+T)=O(N\log(N+2))$ reference realizations on every outcome. The resulting total
query bounds are
\[
 O(N^4)\quad\text{in expectation},\qquad
 O(N^4\log(N+2))\quad\text{in the worst case}.
\]
In particular, when matroid independence queries themselves take polynomial
time, this is a polynomial-time algorithm in the complete rational
input model.

\section{Certifying the general matroid bound}\label{sec:parameters}

The algorithm uses the rational parameters
\begin{equation}\label{eq:parameter-witness}
 \delta=\frac7{4096},\quad \theta=\frac{367878301}{10^9},\quad
 \eta=\frac{247}{5000},\quad s=\frac52,\quad
 \varepsilon=10^{-6},\quad \rho=\varepsilon^3,\quad
 \varepsilon_0=1.14\cdot10^{-6}.
\end{equation}
Two observations allow a slightly better choice of $\theta$.
The threshold step is used only for $|q-1|<3\delta$. Also, although the
curves $a(q)$ and $1-e^{-1/q}$ are nearly tangent at $q=1$, their gap
decreases quadratically away from that point. We retain this quadratic
term when checking the continuation near the middle of $J$.

\begin{lemma}[Parameter choice]\label{lem:parameter-choice}
The parameters in~\eqref{eq:parameter-witness} satisfy
\eqref{eq:basic-parameters} and the three branch conditions
\eqref{eq:condition-guard}--\eqref{eq:condition-threshold}.
Throughout $J$ one has $\omega>0$ and $q(b+s)>1$. Moreover,
$1-\theta-2\rho>1-1/e+\varepsilon_0$.
\end{lemma}

\begin{proof}
Write $u=e^{-1}$. We use the rational enclosures
\[
 u_-:=\frac{36787944117}{10^{11}}<u<
 \frac{36787944118}{10^{11}}=:u_+,\qquad e^{-5/2}<\frac{821}{10000}.
\]
For the first, bound $e$ between $\sum_{j=0}^{16}1/j!$ and that sum plus
$2/17!$. For the second, use the first eleven terms of the positive
series for $e^{5/2}$. The basic restrictions and $q(b+s)>1$ follow
from $0.99<q<1.01$, $0.36<\theta<0.37$, and~\eqref{eq:parameter-witness}.
The final ratio follows from $u_- -\theta-2\rho>1.14\cdot10^{-6}$.

Put $a_0(q)=1-e^{-1/q}$. Its curvature on $J$ satisfies
$a_0''(q)=e^{-1/q}(2q-1)/q^4>1/3$: here
$e^{-1/q}\ge u(1-1/99)>0.36$.
Since $a-a_0$ and its derivative both equal $u-\theta$ at $1$,
\begin{equation}\label{eq:parameter-affine-surplus}
 a(q)-a_0(q)\le q(u-\theta)-(q-1)^2/6.
\end{equation}
Let $x=\lambda\delta$, so $e^{-\lambda}=\alpha e^{-x}$ and
$\alpha/\omega=1+x/\omega$. Applying
$1-e^{-t}\ge t-t^2/2$ at $t=x/\omega$ and
$e^{-x}\ge1-x+x^2/2-x^3/6$ in~\eqref{eq:controller-phi} gives, for $z\ge0$,
\begin{equation}\label{eq:parameter-controller-taylor}
\begin{aligned}
 \Phi_\delta(q,z)-a_0(q)\ge{}
 &-x^2\left(\frac{1-\alpha}{2}+\frac{x}{2\omega}
                             +\frac{\alpha x}{6}\right)\\
 &+ x\alpha(1-\alpha)\left(1-\frac{x}{2\omega}\right)z .
\end{aligned}
\end{equation}

It remains to bound rational expressions on intervals.
Partition $J$ at the points $1+i\delta/32$, $-128\le i\le128$.
For consecutive endpoints $l,r$, put
\[
 A=u_-\left(2-\frac{1-\delta}{l}\right),\quad
 H=\frac{u_+r}{1-\delta},\quad X=\frac{\delta}{l},\quad W=A-X .
\]
The inequalities $1+y\le e^y\le1/(1-y)$ for $y<1$, applied at
$y=1-(1-\delta)/q$, give $A\le\alpha\le H$. Every interval has
$0<A<H<1/2$ and $W>X>0$, so $\omega\ge W>0$.
Since $t(1-t)$ increases on $[0,1/2]$, define
\[
 C=\frac1{l^2}\left(\frac{1-A}{2}+\frac{X}{2W}+\frac{HX}{6}\right),
 \qquad B=\frac{A(1-A)}r\left(1-\frac{X}{2W}\right).
\]
Combining~\eqref{eq:parameter-affine-surplus}
and~\eqref{eq:parameter-controller-taylor}, we obtain
$\Phi_\delta(q,z)-a(q)\ge G+B\delta z$ throughout the interval, with
\[
 G=-C\delta^2-r(u_+-\theta)+
                       \tfrac16\max\{l-1,1-r,0\}^2 .
\]
Also $z_*(q)\ge z_*(l)$ since $d_*$ increases. For a threshold step,
$b$ decreases and $d_*$ increases, so put
\[
 P=d_*(l)+20\varepsilon-1+b(r)(1-821/10000).
\]
Where $P>0$, the left side of~\eqref{eq:condition-threshold} minus its
right side is at least
\[
 T=1-b(l)+\frac{b(r)P}{b(r)+s}
           -b(l)r\bigl(d_*(r)+\eta+30\varepsilon\bigr)-2\varepsilon .
\]
Thus the finite checks in the table suffice. Intervals meeting at a
branch boundary are closed, so their endpoints are covered.
\begin{center}
\begin{tabular}{lrr}
\toprule
Lower bound tested & Intervals & Strict lower bound\\
\midrule
$G$, where $|q-1|\ge3\delta$ & 64 & $2.3\cdot10^{-6}$\\
$G+B\delta z_*(l)$, throughout $J$ & 256 & $7\cdot10^{-10}$\\
$T$, where $|q-1|\le3\delta$ & 192 & $1.13\cdot10^{-5}$\\
\bottomrule
\end{tabular}
\end{center}
\end{proof}

\begin{proof}[Proof of Theorem~\ref{thm:main}]
First consider explicit finite distributions with rational values and probabilities.
Use the parameters in Lemma~\ref{lem:parameter-choice}.
Proposition~\ref{prop:ideal-guarantee} gives the ideal policy
expected reward at least $(1-\theta-\rho)F_0$.
Lemma~\ref{lem:combined-bit-coupling} loses at most another
$\rho F_0$, so the implemented policy obtains
$(1-\theta-2\rho)F_0>(1-1/e+\varepsilon_0)F_0$ when $F_0>0$.
For $F_0=0$ the asserted inequality holds trivially.
The query bounds and worst-case polynomial arithmetic and
random-bit bounds are those of
Appendix~\ref{sec:combined-implementation}.

For general distributions, discard loops and assume $F_0>0$.
Put $\beta=1-\theta-2\rho$ and $c=1-1/e+\varepsilon_0$, so
$\beta>c$. For any $\tau>0$, truncation and downward rounding
to a finite rational grid produce independent $Z_e\le X_e$,
bounded by a rational $B$, with
$\E\OPT(M,Z)\ge F_0-\tau$.
Indeed, the truncated optima increase to the original optimum,
and rounding loses at most $N$ times the grid width.
Approximate each discretized marginal by a positive rational
probability vector on the same support. This gives a product
distribution $Y$ whose joint total variation distance from $Z$ is
at most any prescribed $\gamma>0$. Here the total variation distance
between two distributions $P$ and $Q$ on a finite set is
$\tfrac12\sum_x|P(x)-Q(x)|$; equivalently, it is the minimum
probability that coupled draws from $P$ and $Q$ differ.

Run the finite algorithm designed for $Y$ on the rounded values
$Z$. Couple $Y$ and $Z$ and use the same arrival order and
internal randomness. The two rewards coincide whenever their
value vectors coincide, and both rewards and both offline
optima lie in $[0,NB]$. Hence the expected reward on $Z$ is
at least
\[
 \beta\,\E\OPT(M,Z)-(1+\beta)NB\gamma
 \ge \beta F_0-\beta\tau-(1+\beta)NB\gamma.
\]
At an actual arrival, form $Z_e$ from $X_e$ and feed it to this
policy. The accepted set earns at least its rounded value.
Choose $\tau$ and then $\gamma$ so that
$\beta\tau+(1+\beta)NB\gamma<(\beta-c)F_0$.
This proves the same factor $c$ for general distributions.
Only the explicit finite rational model is used in the
computational claim.
\end{proof}

\end{document}